\documentclass[default,iicol]{sn-jnl}

\usepackage{graphicx}%
\usepackage{multirow}%
\usepackage{amsmath,amssymb,amsfonts}%
\usepackage{amsthm}%
\usepackage{mathrsfs}%
\usepackage[title]{appendix}%
\usepackage{xcolor}%
\usepackage{textcomp}%
\usepackage{manyfoot}%
\usepackage{booktabs}%
\usepackage{algorithm}%
\usepackage{algorithmicx}%
\usepackage{algpseudocode}%
\usepackage{listings}%
\usepackage{natbib}
\setcitestyle{authoryear,open={(},close={)}}
\usepackage{makecell}
\usepackage{color}
\usepackage{pifont}
\usepackage{rotating}
\usepackage{lscape}
\usepackage{tabularx}
\usepackage{longtable}
\usepackage{pdflscape}
\usepackage{url}
\let\citep\cite
\renewcommand{\cite}[1]{(\citeauthor{#1},\citeyear{#1})}

\begin{document}

This work has been submitted to IEEE for possible publication. Copyright may be transferred without notice, after which this version may no longer be accessible.

\newpage

\title[Article Title]{A Review of Prototyping in XR: Linking Extended Reality to Digital Fabrication}


\author[1]{\fnm{Bixun} \sur{Chen}}\email{2429283C@student.gla.ac.uk}

\author[2]{\fnm{Shaun} \sur{Macdonald}}\email{shaun.macdonald@glasgow.ac.uk}

\author[3]{\fnm{Moataz} \sur{Attallah}}\email{m.m.attallah@bham.ac.uk}

\author[4]{\fnm{Paul} \sur{Chapman}}\email{p.chapman@gsa.ac.uk}

\author[1*]{\fnm{Rami} \sur{Ghannam}}\email{rami.ghannam@glasgow.ac.uk}

\affil*[1]{\orgdiv{James Watt School of Engineering}, \orgname{University of Glasgow}, \orgaddress{\city{Glasgow}, \postcode{G12 8QQ}, \country{UK}}}
\affil[2]{\orgdiv{School of Computing Science}, \orgname{University of Glasgow}, \orgaddress{\city{Glasgow}, \postcode{G12 8QQ}, \country{UK}}}
\affil[3]{\orgdiv{School of Metallurgy \& Materials}, \orgname{University of Birmingham}, \orgaddress{\city{Birmingham}, \postcode{B15 2TT}, \country{UK}}}
\affil[4]{\orgdiv{School of Innovation \& Technology}, \orgname{Glasgow School of Art}, \orgaddress{\city{Glasgow}, \postcode{G51 1EA}, \country{UK}}}





\abstract{Extended Reality (XR) has expanded the horizons of entertainment and social life and shows great potential in the manufacturing industry. Prototyping in XR can help designers make initial proposals and iterations at low cost before manufacturers and investors decide whether to invest in research, development or even production. According to the literature (54 manuscripts in the last 15 years) prototyping in XR in XR is easier to use than three-dimensional (3D) modeling with a personal computer and more capable of displaying 3D structures than paper drawing. In this comprehensive review, we systematically surveyed the literature on prototyping in XR and discussed the possibility of transferring created virtual prototypes from XR to commonly used 3D modeling software and reality. We proposed five research questions regarding prototyping in XR. They are: what the constituent elements and workflow of prototyping are; which display devices can deliver satisfying immersive and interactive experiences; how user control input is obtained and what methods are available for users to interact with virtual elements and create XR prototypes; what approaches can facilitate the connection with fabrication to ensure a smooth transition from the virtual to the physical world; and what the challenges are and what the future holds for this research domain. Based on these questions, we summarized the components and workflows of prototyping in XR. Moreover, we present an overview of the latest trends in display device evolution, control technologies, digital model construction, and manufacturing processes. In view of these latest developments and gaps, we speculated on the challenges and opportunities in the field of prototyping in XR, especially in linking extended reality to digital fabrication, with the aim of guiding researchers towards new research directions.}




\maketitle

\section{Introduction}\label{sec1}


Computer-Aided Design (CAD) has undergone a significant transformation since its inception, moving from two dimensions (2D) representations to complex three dimensions (3D) models. Today, it is an integral part of most industrial design processes, enabling designers to visualize, test and iterate upon their prototypes before they reach the physical world. Historically, these processes were confined to developing and visualizing designs using 2D screens, presenting barriers to spatial understanding and user interaction. However, with the advancement of computer hardware and software, Extended Reality (XR) display devices have emerged, offering the ability to be engage with immersive or augmented environments and objects using lifelike interaction \cite{Azuma2001}. 

XR is a term that encapsulates Virtual Reality (VR), Augmented Reality (AR) and Mixed Reality (MR). It  enables designers to visualise, manipulate and experience designs in 3D within a virtual environment \cite{milgram1995augmented}. Traditional virtual prototyping, in contrast, involves creating digital models and simulations to evaluate design concepts, typically leveraging computer-aided design tools in non-immersive environments. Prototyping in XR, however, involves generating interactive, immersive prototypes using XR technologies and even test and visualize certain design aspects in a spatial context. The integration of XR into the design workflow, known as prototyping in XR, has attracted attention and investment, since it promises to transform digital fabrication and design \cite{sherman2018understanding}. For example, users from across the globe can interact with computer-generated environments and other users, fostering real-time collaboration and innovation \cite{carmigniani2011augmented}, which helps reduce development time and costs while enhancing design quality and user experience. 

The surge in popularity of XR for prototyping is driven not only to advancements in hardware but also by the emergence and evolution of software. Blender, a widely used 3D computer graphics software, now provides a version that is ported to the OpenXR platform \cite{OpenXR}. Gravity Sketch \cite{gravityS}, an industry-trusted 3D design and modelling software also provides a version that supports VR Head Mounted Display (HMD) devices. In 2007, Jimeno and Puerta observed the rapid development of virtual reality technology and explored its potential application in industrial design and manufacturing processes \cite{jimeno2007state}. They identified that the devices at that time had limited speed and accuracy in handling 3D application scenarios. Since then the Human-Computer Interaction (HCI) experience and capabilities of XR HMDs are constantly being enriched. The release of Microsoft's Hololens in 2015 demonstrated seamless gesture and eye movement tracking using integrated cameras led to several subsequent off-the-shelf HMDs to incorporate similar functionalities, thereby elevating the user experience in virtual environments. In 2021 Varjo released Varjo XR-3 and Varjo VR-3, which provided 60 angular pixel visibility that was equal to the standard visual acuity of the human eye. Furthermore, various biosensors such as heart rate monitors can be integrated into HMDs, and facial expression tracking can be achieved using head-mounted cameras on consumer VR headsets \cite{hpg2}. Additionally, algorithms for animating facial expressions on avatars are also being developed \cite{bai2024universal}. All these advances in HCI have made 3D digital prototyping in the virtual world a promising design method for the future.

This work provides a detailed review of the depth and breadth of prototyping in XR, examining its historical context, current applications and future potential. We focus on how researchers in the past 15 years addressed user interaction, prototyping methods and and the transition from XR prototyping to physical fabrication.
The review is organised as follows: Section 1 introduced the background and motivations for exploring this topic. In Section 2, we discuss the concept of prototyping in XR, reviewing prior work with a focus on studies that used both head-mounted and non-head-mounted displays for prototyping. In Section 3, we introduce our research objectives, research questions and the methodology for selecting articles to be reviewed. In Section 4, we explores XR technologies used for prototyping in both academic research and across various industries. We then address Research Question 1 (RQ1) by examining the key building blocks and workflows in XR prototyping in Section 5.  Section 6 examines device usage trends to answer RQ2, highlighting the advantages and trade-offs of different XR devices. Section 7 addresses RQ3 by reviewing the control methods used in XR prototyping, including input techniques like mid-air gestures and touch interactions. In Section 8, we explore RQ4 by reviewing the literature that show how XR prototyping informs and supports the path to fabrication, categorising these approaches into manual and machine fabrication. Section 9 answers RQ5 by discussing the benefits, challenges, and future potential of XR prototyping, based on the six core building blocks. Finally, we summarise the key findings of our review in Section 10.

\section{Prototyping in XR}\label{sec2}

In this section, we present the semantic definition of ``Prototyping in XR'', then will present research projects from the literature that illustrate its practical application and potential.

\subsection{XR Prototyping}\label{XR Prototyping}
Prototyping in XR is the process of creating a sample or model through XR display devices to give a visual preview or a printable 3D model that helps the designer to test the design concept and its usability. 

\begin{figure*}
\centering
\includegraphics[width=6in]{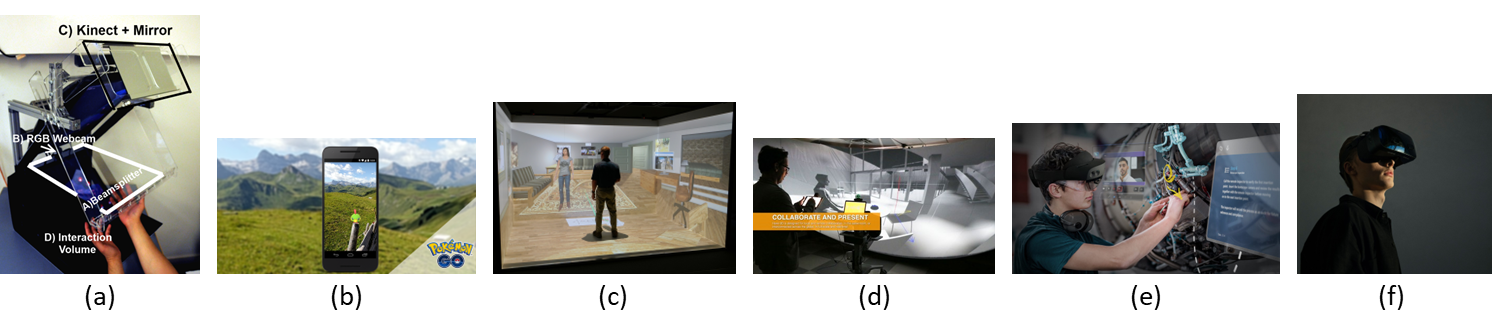}
\caption{Various XR display device, (a) Holodesk \cite{hilliges2012holodesk}, (b) smartphone for AR \cite{poke}, (c) CAVE \cite{cave}, (d) Hybrid Virtual Environment 3D (Hyve-3D) is designed by Hybridlab to facilitate the initial stage of 3D content creation in virtual environments, (e) HMD for AR \cite{bibAR}, and (f) HMD for VR \cite{varjo}.}
\label{fig_sim}       
\end{figure*}

XR display devices are designed to offer users environments that range from fully immersive to partially immersive experiences. Using these devices, users can engage with AR through smartphones or desktop displays, interact with both holographic projections and standard 2D images, or experience video content through HMDs. The application scenes with various XR display devices are demonstrated in Fig. \ref{fig_sim}. We speculate that prototyping in XR serves to provide designers with a more life-like tool for prototyping than traditional CAD software on a computer. The improvement in realism and engagement provided by allowing users to share space and naturally engage with prototyped designs could helps designers and stakeholders better understand their designs, allowing for quick changes and improvements.

From a broader perspective which is not discussed in this paper, XR prototyping can also encompass the process of using XR display devices to create and generate prototypes for various products, including digital artefacts such as films, games and other applications \cite{nebeling2020xrdirector, gruenefeld2022vrception}.
Prototyping in XR could be evaluated through the dimension of prototyping fidelity and virtuality in reality–virtuality continuum \cite{milgram1995augmented, mann2023extended}. This framework allows for assessing the level of immersion and realism in XR prototypes, ranging from low-fidelity sketches to high-fidelity virtual models. Typically, sketching is a form of low-fidelity prototyping offering a basic concept and a rough visual design. While it is quick and flexible, it may lack the detail needed for more advanced testing. Higher-fidelity prototypes, such as detailed 3D models, offer more accuracy but require more resources. Evaluating prototypes based on fidelity and virtuality can guide designers in choosing the right balance between speed and realism for their project.

\subsection{XR Prototyping on Screens}
XR display devices are diverse, with smartphones and monitors that provide a partially immersive experience being most widely available to the public. Cecil Piya and Vinayak proposed RealFusion in 2016 to obtain the 3D model of a physical object with depth camera and enable user to edit 3D models on a monitor to create prototypes \cite{cecil2016realfusion}. This type of workflow for prototyping, which creates digital avatars of physical prototypes for the user to interact with, could be classified as ``physical-based prototyping in XR''. Juggles, clay models and paper drawings are commonly used physical prototypes that can be extended virtually. For example, the ProtoAR and 360proto applications introduced by Nebeling \textit{et al.} \cite{nebeling2018protoar, nebeling2019360proto} can rapidly create virtual prototypes from paper and PlayDoh prototypes with built-in AR capabilities of smartphones.

Various XR display devices are discussed with capacity for prototyping in XR. HoloDesk \cite{hilliges2012holodesk} is a situated see-through display system that allows users to interact with virtual 3D graphics on a desktop surface. Weichel \textit{et al.} used the depth camera to recognize gestures to create virtual objects and introduce existing physical objects into the design based on HoloDesk \cite{weichel2014mixfab}.

\subsection{XR Prototyping in HMDs}
The increase in available computing power of HMDs has facilitated the implementation of prototyping in XR in various products and research works. There are also many academic articles in the literature that have explored the potential benefits of more immersive prototyping in XR methods. For example, the lower learning threshold of manipulation and prototyping is a hot topic \cite{fu2022easyvrmodeling, freitas2020systematic}.
There are many commercial software products that enable users to sketch or build 3D models in the virtual world with HMDs. Google's Tilt Brush \cite{tiltB}, Open Brush \cite{opB}, Microsoft's Microsoft Marquette and Sketchbox's Sketchbox 3D are 3D painting applications that allow users to sketch 3D virtual pen brushes. These applications provide a more naturalistic prototyping experience that is closer to the physical painting process with paper and pen. On the other hand, Google's Google Block \cite{GGBlocks} is a 3D modelling application that enables users to create 3D models in VR in a similar way to traditional CAD, whereby 3D model created by the user in Google Block are regular geometry that can be spliced together, and the shape of the model can be changed by grasping the anchor points/vertices.

These commercial products mostly focus on visual rendering to provide users with a smooth prototyping experience in XR, while researchers are working on creating brand new tools to explore the wider potential of XR prototyping with the new generation HMDs. Situated Modelling, proposed by Lau \textit{at al.}, used marker-attached handles to facilitate AR prototyping \cite{lau2012situated}. The attached markers are recognized either as a geometric overlay on the real-world scene or as a command to generate a series of duplicates along the path of the user's sweeping gesture. To give the capability to bring users' virtual prototypes to the physical world, the shapes that matched with the markers are virtual copies of a set of ready-made wooden blocks. Peng \textit{et al.} introduced a system that allows for 3D models to be designed and 3D printed in almost real time, offering quick physical feedback when the designer is prototyping a 3D object with an AR headset \cite{peng2018roma}.

\section{Methodology}\label{sec3}

In this section, we discuss our research objectives and our methodology for collecting and synthesising the literature on XR for digital prototyping and fabrication. 

\subsection{Research Objectives}

The key objectives of our systematic review article are:

\begin{enumerate}
    \item [O1:] To review the range of current research on prototyping in XR.
    \item [O2:] To provide an overview of the components used for developing virtual prototypes and identify the focal interest points.
\end{enumerate}

\subsection{Research Questions}\label{RQ}
\begin{figure*}[htbp]
\centering
\includegraphics[width=6in]{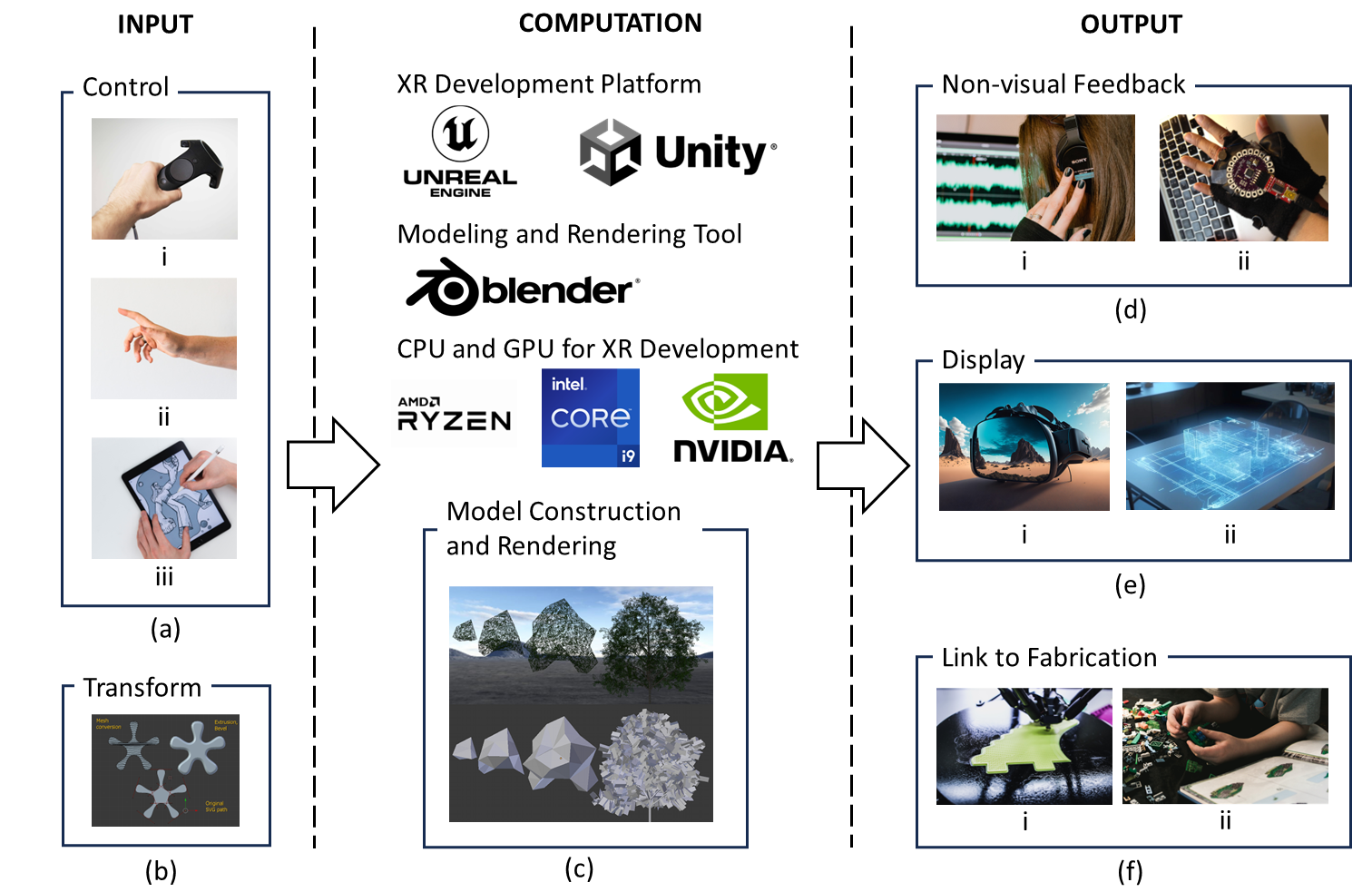}
\caption{Illustration of the six research topics with typical application examples, which include (a) Control: (i)controller, (ii)hand gesture and (iii)touch screen gesture with stylus pen); (b)Transform; (c) Model Construction and Rendering; (d) Non-Visual Feedback: (i)audio and (ii)haptic; (e)Display: (i)VR HMD and VR view and (ii) hologram; and (f) Link to Fabrication: (i) printable files for auto fabrication by 3D printer and (ii) instructions for manual fabrication).}
\label{topic}
\end{figure*}

Based on a preliminary survey, the research area of XR prototyping could be divided into the following six topics, as demonstrated in Fig. \ref{topic}:
\begin{enumerate}
  \item Displays: 
  
  This topic explores the display methods used to achieve partial or fully immersive prototyping experiences.
  \item Control:
  
  This includes the solutions for control and semantics, examining how users interact with and manipulate the virtual environment as well as the implications these have on the design and functionality of XR prototypes. 
  
  \item Model Construction and Rendering:
  
  This area focuses on how virtual prototypes are made and shown, exploring the methods and technologies used in their creation and visualization. 
  
  \item Transform:
  
  Research on this topic examines methods for converting 3D models from traditional CAD to VR prototypes and aims to fill the format gap from the prototyping platform shift.
  
  \item Non-Visual Feedback:
  
  Non-visual feedback such as haptic and olfactory feedback. This research topic explores how these additional sensory inputs can contribute to a more engaging and realistic user experience during prototyping.
  
  \item Link to Fabrication:
  
  This theme focuses on bringing virtual prototypes into the real world in a natural and smooth manner. It aims to combine the advantages of strong immersion and low production costs of prototyping in XR and the intuitive effects of physical prototypes.
\end{enumerate}

Accordingly, our systematic review aims to answer the following five research question (RQs):

\begin{enumerate}
    \item [RQ1:] What are the building blocks and workflow of prototyping? (Are the prototypes built from physical-based reference or totally created on a blank canvas in XR world?)
    \item [RQ2:] Which display devices are capable of delivering satisfying immersive and interactive experiences?
    \item [RQ3:] How is user control input obtained and what methods are available for users to interact with virtual elements and create XR prototypes?
    \item [RQ4:] What approaches can link prototyping in XR with fabrication, ensuring a smooth transition from the virtual to the physical world?
    \item [RQ5:] What are the challenges of prototyping in XR and what does the future hold for this research domain?
\end{enumerate}

\subsection{Review Protocol}
We have set up a review protocol to guide our systematic review on XR prototyping. In this section, we briefly outline our approach, covering our search strategy, inclusion criteria, exclusion criteria and screening mechanisms for selecting relevant research papers.

\subsubsection{Search Strategy}\label{Search Strategy}
Our review considered the latest research articles from major publishers that include IET, Science Direct, Nature, AIP, ACM digital library, Wiley, IEEE Explorer, IoP science, ACS publications and MDPI. Our search also included non-pre-reviewed articles from arXiv. Thus, we performed the critical appraisal using the AACODS (Authority, Accuracy, Coverage, Objectivity, Date, Significance) checklist \cite{tyndall2010aacods} as an evaluation and critical appraisal tool of grey literature (publications and research created by groups not affiliated with conventional academic or commercial publishing institutions).

We begin with querying all the repositories with different research items. As previously mentioned, we put particular focus on XR prototyping and connecting virtual prototypes with fabrication processes, especially using 3D printing for manufacturing. Table \ref{tab:table1} organizes the keywords used in our research, grouped into three categories that highlight distinct aspects of the study focus. The first category, "Mixed reality environments," encompasses terms like \textit{virtual reality (VR)}, \textit{extended reality (XR)}, \textit{augmented reality(AR)}, and others that describe immersive or partially immersive user experiences. The second category, "Virtual object construction," focuses on the process of creating virtual prototypes, with keywords such as \textit{prototyping}, \textit{virtual prototyping}, \textit{authorizing}, and \textit{modelling} illustrating various approaches to expressing designs in XR scene. Lastly, the third category, "Virtual-to-physical transformation," emphasizes the integration of virtual modelling techniques with 3D printing technologies. This category includes terms like \textit{fabrication}, \textit{rapid prototyping}, and \textit{3D printing}, which detail how virtual models are materialized into physical objects either instantly or with some delay.

When conducting searches, these categories are combined using Boolean operators to refine results. For instance, searches combining "Mixed reality environments" AND "Virtual object construction" explore articles on immersive environments and the creation of virtual prototypes. Similarly, searches using "Mixed reality environments" AND "Virtual-to-physical transformation" retrieve works that focus on connecting immersive environments with fabrication processes. Articles were scanned based on their title and abstract, as well as a full-text read of the publications. In addition, we developed search strings using Boolean operators (AND, OR) to connect these keywords. An example or the search strings is: \textit{Title OR Keyword OR Abstract (“virtual” OR “virtual reality” OR “mixed reality” OR “augmented reality” OR “immersive”) AND (“prototyping” OR ”modelling” OR “sketching” OR “authorizing”) AND Year Published(2008-2023).}

\begin{table*}[!t]
\caption{Synonyms and Definitions of Descriptors Used for Search\label{tab:table1}}
\centering
\renewcommand\arraystretch{1.3}
\begin{tabular}{|p{100pt}p{160pt}p{130pt}|}
\hline
Category & Definition & red{Keywords/Terms}\\
\hline
Mixed reality environments & Mixed reality environments provide users with an immersive or partial immersive experience. & Virtual, Virtual Reality (VR), Augmented Reality (AR), Mixed Reality (MR), Extended Reality (XR), Immersive\\
 & & \\
Process of constructing a virtual object & Process of constructing a virtual object (preferably the form of expression in 3D). & Prototyping, Virtual Prototyping, Modelling, Sketching, Authorizing, Designing\\
 & & \\
Virtual-to-physical transformation & Process of transforming a virtual prototype into a physical object. Specifically, it involves integrating 3D printing technology with virtual modelling techniques to materialize models either instantly or with a delay. & Fabrication, Rapid Prototyping, Physical Prototyping, Real-Time 3D Printing, 3D Printing\\
\hline
\end{tabular}
\end{table*}

\subsubsection{Eligibility criteria}
Publications discussed XR prototyping that matched the deﬁnitions and descriptors in  Sect. \ref{XR Prototyping} were considered. More specifically, we used several inclusion and exclusion criteria. The following are the parameters used in the \textbf{inclusion} criteria.

\begin{enumerate}
    \item We included only English-language articles 
    \item We included articles from the past 15 years (since 2008). 
    \item We included articles which are searching results of the query introduced in  Sect. \ref{Search Strategy}.
    \item We included articles involving the interactive fabrications described in Table \ref{tab:table1}.
\end{enumerate}

The following is a list of the \textbf{exclusion} criteria for shortlisting the research papers based on our research objectives and targeted research questions.

\begin{enumerate}
    \item Research articles published in languages other than English. 
    \item Research papers that are not available in full text. 
    \item Editorials, survey reviews, abstracts, and brief papers involving secondary studies are excluded.
    \item Technical report and patent document are excluded.
    \item Articles that did not address the integration of XR with digital prototyping or fabrication.
    \item Articles that are out of scope, which neither construct digital prototypes nor link existing digital prototypes to fabrication.
	\item Articles that have a workflow that do not involve immersive or partial immersive experience.

\end{enumerate}

\begin{figure*}[h]
\centering
\includegraphics[width=6in]{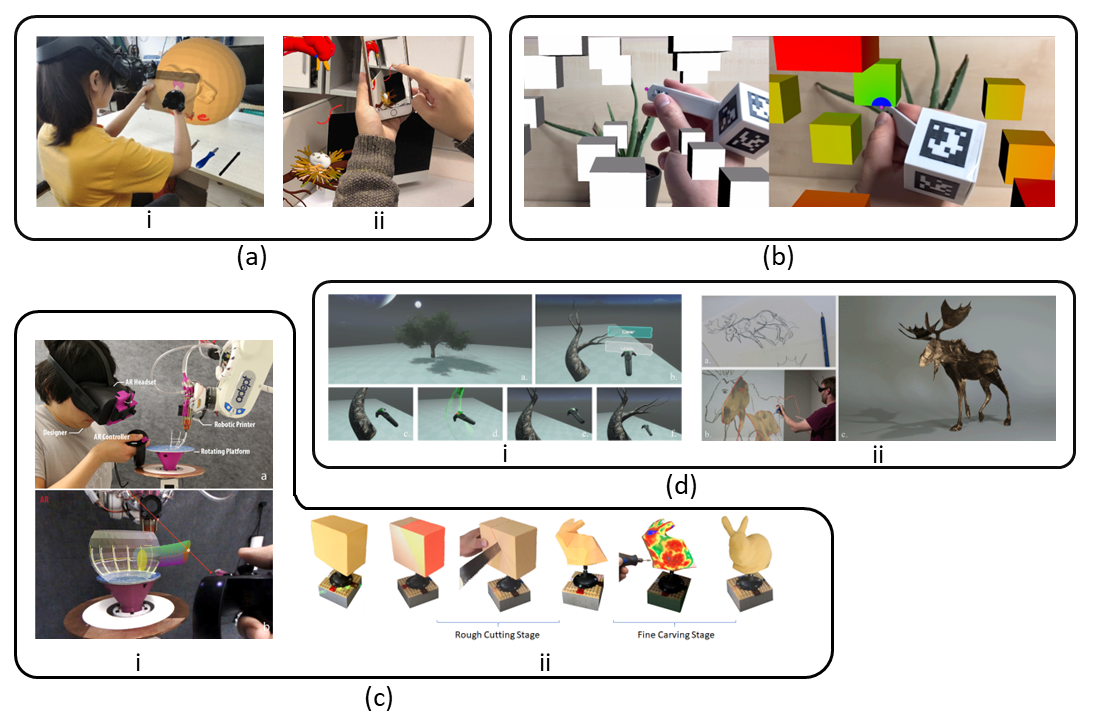}
\caption{Examples of four of the six building blocks (a) Control: (i) Feng \textit{et al.} employed a handwriting pad and pressure-sensitive pen as input devices for carving and relief creation in VR \cite{feng2022pressure}, (ii) Both the surface drawing on phone and the phone's posture and position serve as creative inputs for prototyping \cite{kwan2019mobi3dsketch}. (b) Display: Indicator bubbles utilized to address depth perception limitations in VR displays. \cite{wacker2020heatmaps}. (c) Optional port: (i) RoMA that allows for almost simultaneous prototyping and fabrication \cite{peng2018roma}, (ii) Convert the carving steps calculated from the digital model into projections to provide visual guidance \cite{hattab2019rough}. And (d) Render: (i) The branch shaped brush specifically designed for creating tree prototypes \cite{yuan2021immersive}, (ii) Liftoff for rendering complex and exquisite surfaces with a imported 2D sketched and 3D pen sweeping \cite{jackson2016lift}. }
\label{block}
\end{figure*}

Articles were further screened in two stages. We first checked the title and abstract of each research article retrieved using the aforementioned search string to identify whether it met the inclusion criteria but was not included in the exclusion criteria. We then further screened our articles based on their full-text content. A total of 54 manuscripts satisfied our search criteria.


\section{XR Prototyping Application Areas}\label{sec4}

XR technologies for prototyping has been explored in both academic research and in digital prototyping and manufacturing across various industries.


\subsection{Research applications}
\textbf{Automotive industry:} Researchers have utilized XR technologies to design and prototype both car exteriors and interior lighting systems. For example, Kim \textit{et al.} demonstrated how XR tools could be employed to refine the aesthetics and functionality of vehicle exteriors while enabling real time adjustments to interior lighting configurations \cite{kim2022sketching}. These advancements allow designers to visualize and iterate on complex designs with greater flexibility and efficiency compared to traditional methods.
\newline
\textbf{Interior decoration:} XR has proven valuable for conceptualizing and refining spatial designs. Studies by Park \cite{park2011ar} and Horst \textit{et al.} \cite{horst2020bite}  have shown how XR can assist in designing and prototyping interior spaces, providing immersive visualizations that help designers and clients better understand the spatial relationships and aesthetic choices in real-time.
\newline
\textbf{Education:}
RealitySketch \cite{suzuki2020realitysketch} demonstrates the potential of prototyping in XR for education, allowing users to draw graphics on a mobile AR screen and bind them to physical objects in real time. This dynamic and responsive interaction can be applied to introducing the Classical Mechanics Model in a physical class, enabling students to visualize and interact with concepts intuitively.
\newline
\textbf{Digital Sculpting:} Eroglu \textit{et al.} introduce a groundbreaking virtual creative environment that bridges traditional art forms with modern XR tools. Their system seamlessly transforms 2D images into volumetric 3D objects, allowing artists to extract artistic elements from input materials using VR-based segmentation tools. Relief is then performed interactively by blending height maps that are automatically generated based on the structure and appearance of the input image. The prototype demonstrates how this tool can integrate analog and virtual art workflows, combining the expressive power of traditional painting and sculpting with the creative possibilities of spatial arrangement in VR.

\subsection{Industry applications}


    \textbf{Automotive industry:} VR and AR have been used to create virtual prototypes of car designs, allowing designers and engineers to visualize and interact with the designs in a 3D environment. This helps identify design issues and make improvements before physical prototypes are built. For example, both Ford and Honda used VR to design and evaluate vehicle prototypes \cite{ford, honda}.
    \newline
    \textbf{Architecture and construction:} Architects and engineers use VR and AR to facilitate Building Information Modelling (BIM)\cite{getuli_bim-based_2020}, creating digital models of buildings and infrastructure projects, which can be explored and modified in real-time. This allows stakeholders to visualize the projects and make informed decisions regarding design and construction \cite{schiavi2022bim}.
    \newline
    \textbf{Medical device prototyping:} VR and AR can be used to develop and test the design of medical devices, such as surgical instruments \cite{kordass2002virtual} and implants\cite{monaghesh2023application}, in a virtual environment. 
    This enables faster iterations and reduces the need for physical prototypes, saving time and resources.
    \newline
    \textbf{Aerospace industry:} VR and AR can be used for prototyping in XR of aircraft components and systems \cite{moerland2021application}. Designers and engineers can collaborate and interact with these virtual models to identify design flaws and make improvements.
    \newline
    \textbf{Fashion and apparel:} VR and AR enable fashion designers to create virtual prototypes of garments and accessories \cite{anta}, allowing for faster design iterations and reducing the need for physical samples. For example, VR allows designers to visualize and interact with 3D models of clothing in a fully immersive environment, where they can modify textures, colors, and shapes in real time. The footwear design studio Khamis Studio uses Gravity Sketch to visualize and interact with 3D models of sneakers in a fully immersive environment \cite{GSsneaker}, where the designers can modify textures, colors, and shapes in real time.This combination of technologies not only accelerates the design process but also helps designers make more informed decisions before producing physical samples, thus saving time and resources.



\section{Building Blocks and Workflow}\label{sec5}

We examined the reviewed manuscripts to address RQ1,  establishing the core building blocks and workflows relevant to XR prototyping. 
In particular, we examined if prototypes were predominantly built from physical-based references or created on a blank canvas within XR.

\subsection{Building Blocks}

\begin{figure*}
\centering
\includegraphics[width=6in]{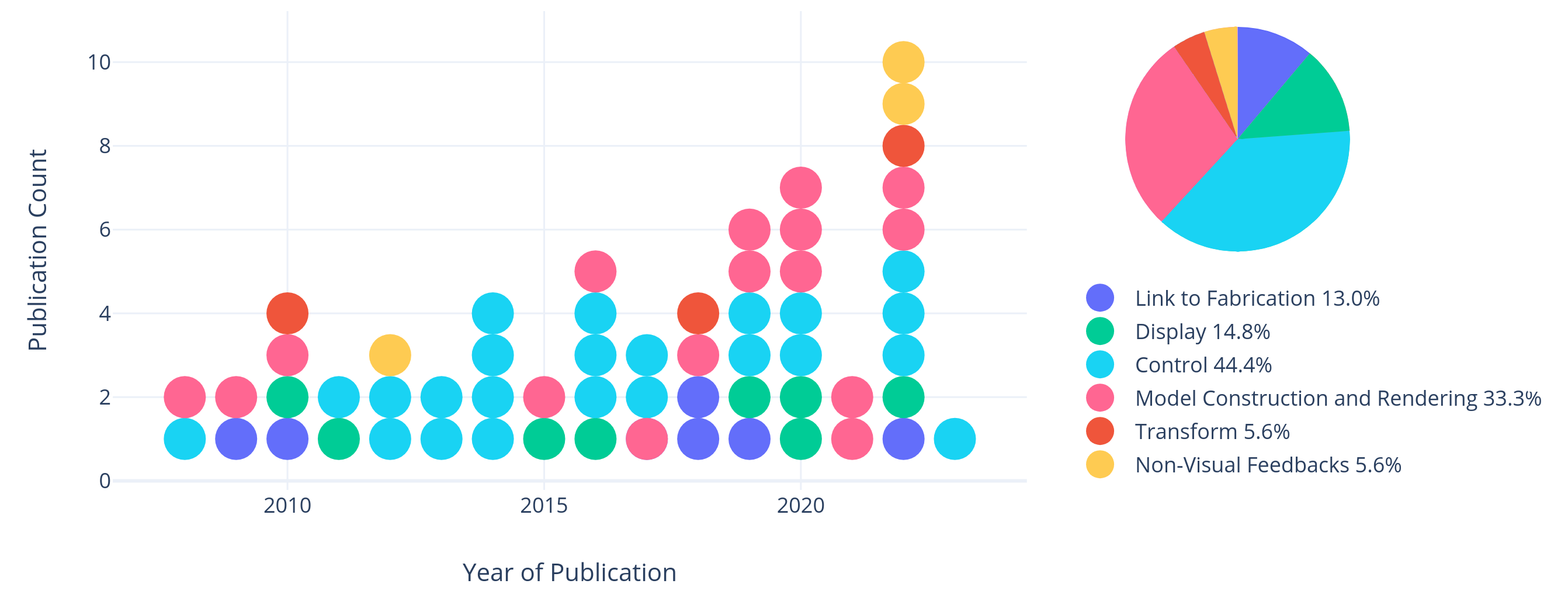}
\caption{Scatter chart depicting publication counts by year and research topic using bubbles and pie chart depicting percentage in each topic.}
\label{line}
\end{figure*}

The articles were assigned to one or more of six categories established in the preliminary survey as outlined in Section \ref{RQ} : (1) display, (2) control, (3) transform, (4) model construction and rendering, (5) non-visual feedback and (6) link to fabrication. The distribution of manuscripts across these categories is depicted in the scatter chart Fig. \ref{line} based on their publication years.

\noindent\textbf{Display: }Approximately 14.8\% of these articles introduce various "Display" methods suited for better visualizing and aiding in the comprehension of prototypes or aimed at achieving either partial or fully immersive prototyping experiences. This encompassed both display devices (hardware) and display software methods like using visual guidance such as bubbles, heatmap, and scaffolding surfaces in UI (User Interface) design (software) as shown in Fig. \ref{block}(b) \cite{xu2023gesturesurface, baerentzen2019signifier}.

\noindent\textbf{Control: }Around 44.4\% of the articles engaged with the topic of ``Control", which encompasses user input through various control techniques, such as tracked stylus pen \cite{feng2022pressure} and touchscreen input \cite{kwan2019mobi3dsketch} demonstrated in Fig. \ref{block}(a). 
In the early stage of XR prototype development, it relied on traditional control devices such as keyboards, mice, and touch screens. The operation was simple but could not meet the complex interaction requirements of prototyping in 3D space.
With the advancement of technology, precise control devices such as tracker pens and VR controllers can input three-dimensional designs in real time with high precision. Another example is that touch screens can intuitively control virtual objects with gestures to improve interaction efficiency.
The current research focuses on multimodal interaction, such as the combination of static gestures, dynamic gestures, and controllers, so that prototyping in XR scenarios can be more accurate, free, and more immersive.

\noindent\textbf{Model Construction and Rendering: }``Model Construction and Rendering'' encompasses the processes of constructing models based on user input and the rendering of computer graphics, accounting for 33.3\% of the articles. Two notable examples are presented in Fig. \ref{block}(d): (i) a specialized branch-shaped brush for creating tree prototypes, which integrates specific design considerations for natural forms \cite{yuan2021immersive}; and (ii) Liftoff, a technique for rendering complex surfaces by combining imported 2D sketches with 3D pen sweeping, facilitating detailed surface creation \cite{jackson2016lift}.

\noindent\textbf{Transform: }Manuscripts were assigned to the ``Transform" category if they proposed or discussed methods for converting 3D models from traditional CAD to VR prototyping, bridging a format gap, with a representation of 5.6\%. Lorenz \textit{et al.} \cite{lorenz2016cad}utilized the VRML (Virtual Reality Modeling Language) standard to facilitate the conversion from CAD models to VR environments, enabling the automatic generation of VR models from CAD animations. However, this approach is limited to Instant Reality, a web-based 3D VR application. In Kim \textit{et al.’}s Cyber Physical System server for VR engineering, a VR Parser was developed that can generate BOM (Bill of Material)-based 3D graphics models as objects in the VR environment for HMDs based on the input CAD files \cite{kim2022design}.

\noindent\textbf{Non-Visual Feedback: }The ``Non-Visual Feedback'' block, assigned to 5.6\% of articles, refers to auxiliary feedback for prototyping, such as reminder tones \cite{fechter2022comparative, xu2022applying}.

\noindent\textbf{Link to Fabrication: }13\% of the manuscripts discussed or explored ``Link to Fabrication" refers to exporting the results of XR prototyping for rapid and on-demand 3D printing or other manufacturing. Two examples of exporting prototyping results as manufacturing instructions are demonstrated in Fig. \ref{block}(c): (i) RoMA, which enables almost simultaneous prototyping and fabrication by integrating design and manufacturing workflows \cite{peng2018roma}; and (ii) a method to convert carving steps from a digital model into projections, providing visual guidance during the fabrication process \cite{hattab2019rough}. 

\subsection{Workflows}

\begin{figure*}[h]
\centering
\includegraphics[width=6in]{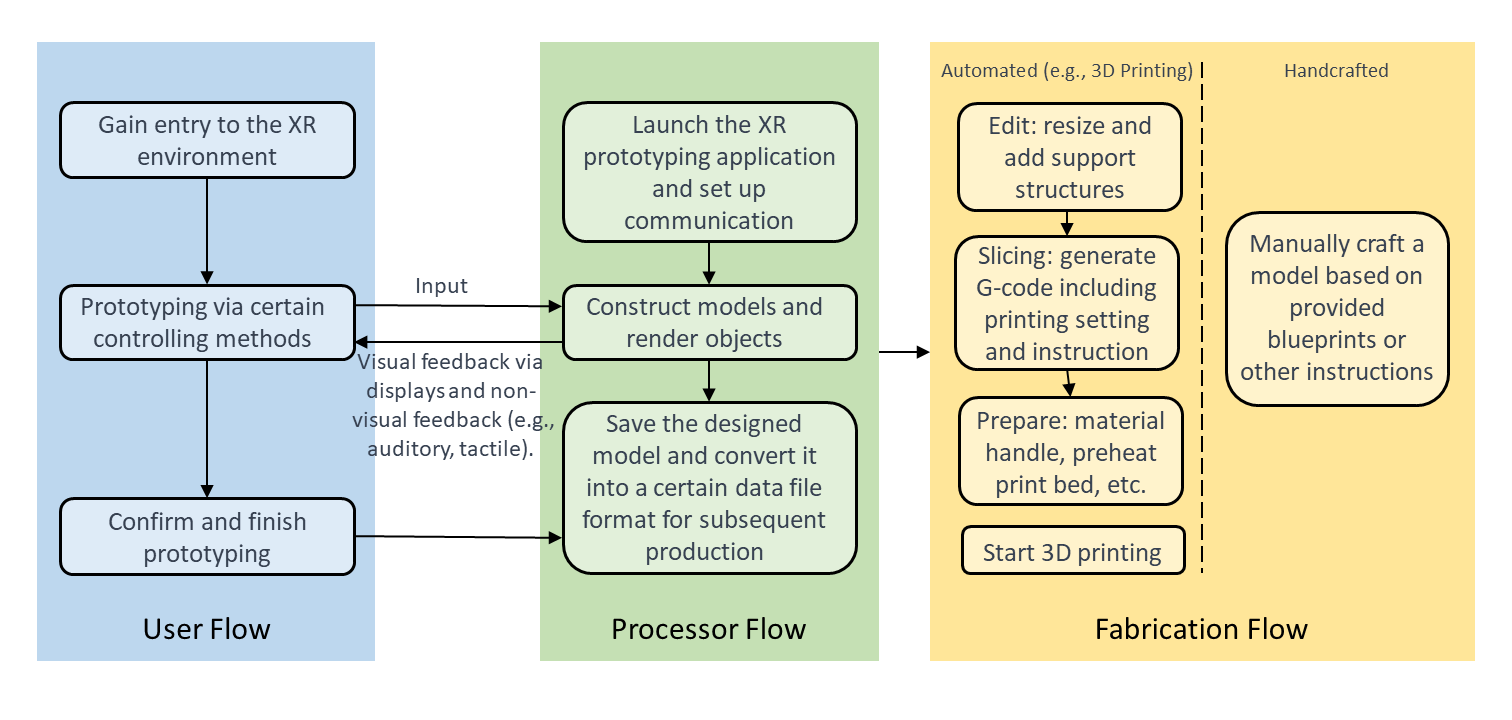}
\caption{Workflow of prototyping within a metaverse environment and its interconnected fabrication process}
\label{workflow}
\end{figure*}

After reviewing the selected articles, we have outlined the abstract workflow for XR prototyping and linking to fabrication, as illustrated in Fig. \ref{workflow}.

\begin{figure*}[h]
\centering
\includegraphics[width=5.3in]{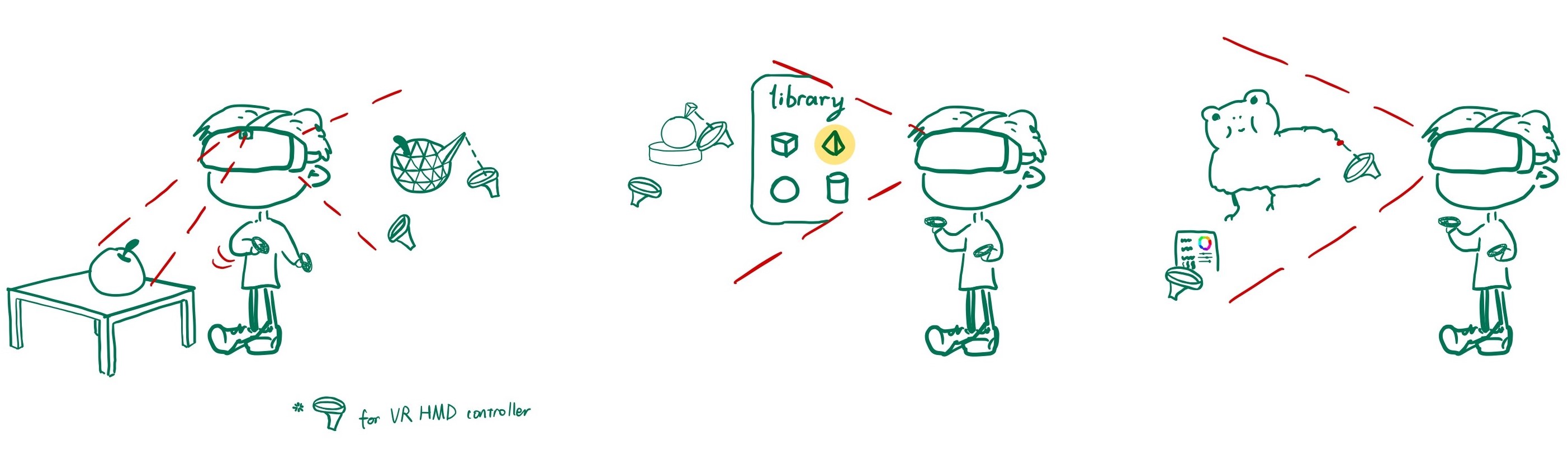}
\caption{Sketch map for the three prototyping methods with a VR HMD and paired controller: physical-based construction and rendering (left), rapid construction by the assembly of preset virtual model blocks (middle) and prototype on a blank virtual canvas (right).}
\label{fig_3p}
\end{figure*}

It became clear that there is no consistent answer to the question: ``Are the prototypes built from physical-based reference or totally created on a blank canvas in XR world?''. Therefore, we organised the literature into three distinct categories, as shown in Fig \ref{fig_3p}. These are: (1) Physical-based construction and rendering, (2) Rapid construction by the assembly of preset virtual model blocks and (3) Prototype on a blank virtual canvas.

\subsubsection{Physical-Based Construction and Rendering}

Physical-based construction and rendering is a method that reconstructs the 3D models from physical objects, then allows the user to reshape, paint or do other operations to customise their own prototypes as shown on the left side of Fig.\ref{fig_3p}.

These physical representations could be 2D paper prototypes or 3D sculptures.
These are works in which 2D paper artefacts or 3D physical models or sculptures are incorporated into the XR world as a creative scaffold, preserving the consistency of designers' creative inspiration or initial reference. 
However, it also increases the complexity of the virtual creative process, as physical templates need to be created and imported first for initial creation. 
\par
Notable examples of this approach include Eroglu \textit{et al.} who introduced their model construction workflow, which reconstructs a sculpting piece model in VR based on the shallow relief work and allows users to modify and reshape based on the sculpting \cite{eroglu2020rilievo}. Jackson \textit{et al.} introduce their system of lifting the curves of the manuscript in virtual space and drawing surfaces to construct prototypes \cite{jackson2016lift}. Marner and their colleagues designed a system that simulates spray painting using a handheld controller and alters the appearance of physical objects by projecting light\cite{marner2011large}. Others such as Huo \textit{et al. used} surface images of physical objects as references for adding textures to virtual models, drawing inspiration from the real world \cite{huo2017window}.
\par
The process of creating virtual 3D content from physical painting was a subtheme of several research manuscripts \cite{bergig2009place, hagbi2010place, hagbi2015place}. For example, work by Habgi \textit{et al.} introduced prototyping systems that interpret physical painting as constructing commands to create 3D content for augmentation according to a predefined visual language.
\par
Physical-based construction and rendering does impose a core limitation on the possibility space of 3D prototyping, as it is reliant on the physical constraints of the subject of its visual input. 
This approach is, however, ideal for creating targeted prototypes. RealitySketch \cite{suzuki2020realitysketch} serves as an example of physical-based construction and rendering, where 3D models are reconstructed from physical objects by binding drawn graphics to them in real time, allowing for dynamic interaction and visualization.\\

\subsubsection{Rapid Construction by the Assembly of Preset Virtual Model Blocks}

Another common approach was to design or utilise preset XR prototyping blocks, allowing efficient modular prototyping with standardised aesthetic or functional components as shown in the middle of Fig.\ref{fig_3p}.
Transformation operations such as translation, scaling and rotation, along with boolean operations like union, intersect and subtract facilitate swift assembly, as highlighted by Fu \textit{et al.} \cite{fu2022easyvrmodeling}. 
This technique is prevalent in projects aiming to swiftly provide users with a preview of personalised products using established components, such as interior decorations \cite{horst2020bite, park2011ar}, car exteriors and car interior lighting \cite{kim2022sketching}.

From an educational perspective, the translation of physical teaching props into virtual reality has been explored to support more sustainable and immersive education. This approach enables students to understand object composition intuitively and even design new component assembly methods for their prototypes. For example, Abriata \textit{et al. applied this approach to create} AR molecular chemistry visualization and modelling kits designed to replace physical plastic modelling, where macromolecular models can be prototyped by combining loaded models from a library \cite{abriata2020building}. 

While this prototyping method is efficient, the reliance on preset blocks may limit customization and creativity of prototypes. Moreover, this method requires a comprehensive library of preset blocks to cater to diverse design needs, potentially limiting its applicability. 
To address this limitation research has explored a variation of this prototyping method where users do not directly select a preset model block in XR by incorporating physical-based construction. 
Instead, they provide a semantic definition for the 2D icons, which are then represented as signs or markers sketched on whiteboards \cite{kim2022cor}, or as drawings or stickers on paper that are converted into 3D XR objects through designed algorithms \cite{nebeling2019360proto}. With the created virtual model blocks, users are then allowed to perform assembly. This method introduces an additional layer of interaction, offering a blend of physical and digital engagement in the prototyping process, while also leveraging the advantage of using pre-established notation and building blocks.\\

\subsubsection{Prototype on a Blank Virtual Canvas}

Prototyping on a blank virtual canvas provides the freest experience, allowing users to create their own prototypes totally according to their idea without the limitation of starting from a reset model or a virtual avatar of a physical object as shown on the right side of Fig.\ref{fig_3p. However, this method also requires more effort in customizing strokes and controlling the rendering algorithms, as the design process demands greater precision and flexibility.} The intuitive method for XR sketching in a blank virtual canvas is to track the trajectory of the controller to create wireframes in XR \cite{lakatos2014t, wacker2019arpen, kwan2019mobi3dsketch, xu2022applying}. The conceptual modeling system CASSIE by Yu \textit{et al.} adopts this model-building method and focuses on optimizing the connection of hand-drawn curves to help users achieve continuous rendering results without the need for continuous operations \cite{yu2021cassie}.

Compared to drawing 2D or 3D wireframes on a blank canvas, drawing solid models is more complex, but can produce more refined creative effects.  A common construction and rendering workflow involves users choosing or sketching a 2D shape and then extruded along the path of the user’s controller \cite{drey2020vrsketchin}.
Additional modifications can be made to the models by performing scaling, cutting, rotating, and boolean operations.
It is also possible to modify models by adjusting the positions of individual mesh vertices\cite{GoogleBlocks, teng2017augmented}. Another common construction and rendering workflow is users directly perform 3D sketching and view the rendered result of the sketched convex shape \cite{wibowo2012dressup, MicroMa}.

In terms of how this approach has been applied, some researchers have focused on creating detailed complex 3D model elements drawn from simple digital sketching or gestures by users. For example, one interesting area by Yuan et al. applied this technique to the XR prototyping of tree/forestry modelling, converting sketched curves drawn by users in VR into a tree with a natural-looking trunk and branches \cite{yuan2021immersive}. Unlike the invoking and deployment mentioned in the Sect. \ref{RQ}, it constructs a shape-matching model in the background based on the user's input and presents the rendering effect. Two other examples include LifeBrush, an application for drawing molecular models in VR along the path of user brush-strokes \cite{davison2019lifebrush} and a hair modelling system developed by Xing \textit{et al.}, which implemented the creation of various hairstyles along user strokes \cite{xing2019hairbrush}.
Furthermore, research has explored variants of this 3D drawing approach, such as Arora and Singh implemented anchored user stroke input in mid-air onto a 3D surface, allowing users to draw patterns on existing 3D model surfaces in VR scenes \cite{arora2021mid}.
Each XR prototyping workflow has its strengths: the blank canvas method offers flexibility but requires more precision, while block-based modeling provides more structure and detail. For manufacturing, such as in automotive and aircraft design, the block-based approach may be more suitable for creating accurate, functional prototypes.

\section{Display Devices For XR Prototyping}\label{sec5}

In order to address RQ2, we examined trends in device usage across the review manuscripts to identify key advantages and trade-offs. As discussed in Section~\ref{sec2}, XR prototyping can be performed using screens, HMDs, and projection imaging devices. The development of prototyping in XR follows industrial hardware development. The display devices used to provide an immersive experience have improved step by step from projectors to holographic projectors, computer monitor to hand-held smartphones and now HMDs with integrated sensors and cameras. 
\par
Across to the reviewed literature, screens were used as the primary display modality for XR prototyping 43.1\% of the articles. Projectors were used in 13.7\% of the articles, while HMDs appear were most prominent, used in 47.1\% of articles. Articles using multiple media are counted in each relevant category. These results suggest that screens and HMDs have been the dominant choices for providing visual feedback in virtual creation over the past 15 years. The median publication year for screen-based articles is 2017.5, while HMD-based articles have a median of 2019. This highlights a growing trend towards HMD use in recent years.

\subsection{Screen-Based Displays}
Screen-based displays include standard monitors, which can be used for basic VR experiences, albeit with a low level of immersion due to the lack of stereoscopic depth and head-tracking capabilities. When paired with additional hardware, such as webcams, monitors can also serve as AR display devices. In research on prototyping in XR, discussions on displays focus more on what to choose and how to use them, rather than on the development of hardware itself, such as improving resolution, reducing latency, and addressing issues like motion sickness. Visual tools, such as bubbles and heatmaps, are used to assist users in understanding complex content, and UI design provides dynamic support for interactions, such as progressive displays. Nowadays, as hardware matures, the focus of displays for prototyping in XR has shifted, with the boundaries between hardware and software becoming blurred. Research is now more focused on fully leveraging the advantages of both hardware and software, using them in synergy to enhance immersion and usability.

Webcam-based AR applications leverage real-time video capture to overlay digital artifacts, creating an augmented reality experience. However, such applications rely heavily on the development of AR toolkits (e.g., ARToolKit), which handle tasks like marker recognition, spatial alignment, and artifact rendering. Advances in these toolkits, along with the introduction of depth cameras, have made AR systems increasingly accessible on screen-based devices, including smartphones and tablets. Platforms such as ARCore and ARKit now allow users to create and interact with digital models directly within the context of their physical environment.

The monitor of a computer with peripheral web cameras is a primitive display device providing a low immersive experience. The webcam provides live action video capture feed, which can then be overlaid with digital artefacts to create an augmented reality image. However, such applications rely heavily on the development of AR toolkits (e.g., ARToolKit), which handle tasks like marker recognition, spatial alignment, and artifact rendering \cite{abriata2020building}. Advances in these toolkits, along with the introduction of depth cameras, have made AR systems increasingly accessible on screen-based devices, including smartphones and tablets. Platforms such as ARCore \cite{ARC} and ARKit \cite{ARK} now create and display digital models on the screen shown within the context of their real environment. 

\subsection{CAVE Projection}

Projection imaging is the display method using projectors to show imaging on flat or curved surfaces. CAVE (Cave Automatic Virtual Environment), is a type of immersive virtual reality environment where projectors are directed to between three and six of the walls of a room-sized cube \cite{cruz1992cave}. The user typically wears stereoscopic glasses to see 3D images projected onto the walls, floor and sometimes the ceiling of the room. By tracking the user's head and adjusting the images projected in real-time based on their perspective and position, the CAVE creates the illusion that the user is fully immersed in a virtual world. This environment allows for a high level of interaction and engagement, making it useful for a variety of applications including scientific visualisation, engineering and interactive art. A hybrid environment integrating a CAVE and a GeoWall, as described in \cite{chen2011hybrid}, demonstrates effective interaction techniques for virtual environments, enabling architects to quickly model building masses with physics-based manipulation and table-prop tools. In contrast, Jackson's work \cite{jackson2016lift} employs a 4-wall CAVE environment with lightweight tools and natural-feeling interactions, enabling intuitive 3D modeling through 2D drawing references, particularly excelling in interactive art applications. A core advantage of CAVEs is that users can experiences an immersive projection without needing to personally engage with a screen interface or HMD, allowing for more naturalistic traversal and presence in the space.

\subsection{HMDs}

HMDs have become the most prominent display type for immersive experiences and are now integrated with various sensors such as a gyroscope, accelerometer, magnetometer, face camera for eye tracking with pupillometry, heart rate sensor and so on which can track user actions and state while interacting with immersive spaces or prototypes. 
Table \ref{tab2} summarizes the functions and features of the HMDs features in the articles we reviewed, including entry-level devices such as the Meta Quest 2 and Meta Quest 3, as well as higher-end alternatives performance such as the HP Reverb G2 Omnicept, Varjo XR-4 and Apple Vision Pro.
The table lists which HMDs are equipped with the following functions: hand tracking, body movement tracking, eye tracking, facial movement, voice command, heart rate monitoring, real - time environment capture, and spatial depth perception. If a device requires an accessory to use a certain function it will be noted as 'accessory needed'. For example, HTC VIVE Pro 2 can achieve independent PC VR by obtaining the official wireless adapter on the shelf. 

\subsubsection{HMD Device Specifications}

The Meta series are characterized by providing an entry-level XR experience with an all-in-one design which tracks movement and the real-world via onboard camera. It includes VR, pass-through AR and the spatial anchor that anchors virtual objects in the real environment.
By contrast, HTC VIVE Pro 2's movement tracking is achieved through external base stations, making it more precise than the in out tracking headsets of the same period. 
Hololens 2 is an AR-specific device which provides optical AR projection on a transparent eyepiece, which can maintain a wide actual field of view to the outside world while being worn.Varjo XR-3 and Varjo XR-4 both provide precise depth awareness to achieve pixel perfect real-time occlusion on the real world with virtual content and digital 3D reconstruction of physical objects, while their precise inside out tracking and 'human eye like' visual bring the current ultimate immersive visual experience; The most attractive feature of HP Reverb G2 Omnicept is that its sensors and algorithms can recognize gaze, pupil position, pupil dilation, eye opening, and heart rate, thus enabling cognitive load assessment and greatly facilitating researchers to quantify the cognition of headset users. 
The biggest feature of Apple Vision Pro is its Apple ecosystem friendliness and suitability for collaborative scenarios.

\subsubsection{HMD Device Prevalence}

Among all the articles collected that use HMD, the most popular commercial head display device are HTC Vive series, including HTC VIVE \cite{davison2019lifebrush, yuan2021immersive, zhu2022sensor, fu2022easyvrmodeling}, HTC VIVE Pro \cite{arora2021mid}, HTC VIVE Pro Eye \cite{xu2023gesturesurface}. It is reasonable to believe that the reason is that in the application scenarios of prototype production, HTC Vive has comprehensive capabilities in terms of price, head display processing ability, developer friendliness of the ecosystem, and positioning and tracking system technology for controlling inputs. 

\begin{table*}
\centering
\rotatebox{90}{
\begin{tabularx}{\textheight}
{@{\extracolsep\fill}p{0.08\textheight} p{0.06\textheight} p{0.061\textheight} p{0.05\textheight} p{0.05\textheight} p{0.05\textheight} p{0.045\textheight} p{0.05\textheight} p{0.068\textheight} p{0.01\textheight} p{0.31\textheight}}

\hline
& & \multicolumn{6}{c}{Human understanding} & \multicolumn{2}{c}{Env. understanding} \\ \cmidrule(lr){3-8}\cmidrule(lr){9-10}
Device & Stand-alone headset & Hand tracking & Body movement tracking & Eye tracking & Facial movement  & Voice command & Heart rate & Real-time environment capture & spatial depth & Key feature\\\hline

Meta  Quest 2 & \ding{52} & \ding{52} & \ding{55} & \ding{55} & \ding{55} & \ding{52} & \ding{55} & \ding{55} & \ding{55} & VR; Hand-tracking \newline and Controllers; Social VR \\  \hline

Meta Quest 3 & \ding{52} & \ding{52} & \ding{52} & \ding{55} & \ding{55} & \ding{52} & \ding{55} & \ding{52}  & \ding{52} & MR; Spatial Anchoring\\  \hline

Meta Quest Pro  & \ding{52} & \ding{52} & \ding{55} & \ding{52} & \ding{52} & \ding{52} & \ding{55} & \ding{52} & \ding{52} &  MR; Advanced Facial–track-\newline ing; Enhanced Pass-through\\ \hline

HTC VIVE  Pro 2 & \ding{119}   & In beta & \ding{55} & \ding{52} & \ding{119} & \ding{52}  & \ding{55} & low revolution  & \ding{55} & VR; Precise movement \newline tracking (with base stations) \\  \hline

Microsoft Hololens  2 & \ding{52} & \ding{52} & \ding{55} & \ding{52} & \ding{55} & \ding{52} & \ding{55} & Optical AR & \ding{52} & AR; Mixed reality for pro-\newline ductivity; Enterprise-grade \newline security\\  \hline

Varjo  XR-3 & \ding{55} & \ding{52} & \ding{55} & \ding{52} & \ding{55} & \ding{55}& \ding{55} & \ding{52} & \ding{52} & MR; Depth awareness; 'Hu-\newline man-eye-like' visuals; Precise \newline hand tracking; Enterprise-\newline grade security\\  \hline

HP Reverb G2 Omnicept & \ding{55} & \ding{55} & Arm tracking  & \ding{52} & \ding{52} & \ding{55}   & \ding{52} & \ding{55} & \ding{55} & VR; Cognitive load \newline assessment \\  \hline

Varjo XR-4 & \ding{55} & \ding{119} & \ding{55} & \ding{52} & \ding{55} & \ding{52} & \ding{55} & \ding{52} & \ding{52} & MR; Depth awareness; 'Hu-\newline man-eye-like' visuals; Precise \newline movement tracking; Enter-\newline prise-grade security\\ \hline

Apple  Vision Pro & \ding{52} & \ding{52} & \ding{55} & \ding{52} & \ding{55} & \ding{52} & \ding{55} & \ding{52} & \ding{52} & \raggedright MR; Multimodal interaction;\newline Integration with Apple \newline ecosystem; 'Human-eye-like' \newline visuals \\  \hline      
\end{tabularx}
}
\caption{The technical specification and key features of the mentioned headsets in surveyed papers as well as the most advanced headsets.}
\label{tab2}

\end{table*}

\subsection{Display Considerations for XR Prototyping}

When prototyping in XR, several crucial factors should be considered when selecting display devices: the capability to precisely present intricate and detailed designs; the capability to fulfil users' demand for \textbf{immersive experiences}; and the \textbf{cost-effectiveness} of the devices. Using screens as a medium is advantageous for precise content creation and cost-effectiveness, but immersive experiences are greatly limited by the display format of field-of-view and single-viewpoint imagery. Headsets have greater advantages in terms of immersion, providing realistic stereoscopic content and a sense of presence \cite{mcgill2022augmented}. However, researchers like Chang \textit{et al.} have highlighted negative aspects of HMD use, such as the weight of the device or cybersickness \cite{chang2020virtual}. Saredakis \textit{et al.}'s review article further stresses that continuous exposure to VR gaming content for over 10 minutes or simple VR scenes (such as landscapes) for over 20 minutes using a headset can lead to significant cybersickness \cite{saredakis2020factors}, hampering the potential for longer prototyping sessions.
\par
Additionally, HMDs face greater challenges in creating high-fidelity models due to depth perception \cite{el2019survey}. Projection display media such as the CAVE can provide room-scale immersive environments allowing multiple users to intuitively collaborate on creation. However, they also face issues with unclear three-dimensional depth perception, which limits interaction precision. Considering the cost of devices, it should be noted that our review does not consider the price of computers used for development, but focuses on comparing peripheral device prices. Generally, the price of projectors is higher than that of HMDs, which may be higher than that of screen devices \cite{VRcompare, gamingmonitor}. Therefore, if the designer aims to create highly detailed content and require users to engage in continuous creation for extended periods (\textgreater20 minutes) without emphasising high immersion and life-like stereoscopic presentation, screen devices such as monitors are recommended as a pricier and higher quality choice. If seeking immersion and interactivity, desiring to allow local collaborative creation, while not requiring very high levels of detail in the content, then using projectors could be considered. Meanwhile, projection display media like CAVE systems may involve higher initial setup costs but could provide cost savings in the long run by accommodating multiple users in collaborative environments without the need for individual headsets. Conversely, if aiming to provide users with a superior immersive experience using a larger field-of-view, stereoscopic rendering capabilities, and integrated sensors, and only needing to create conceptual models or other content with low requirements for accuracy and precision, without requiring prolonged user engagement, HMDs are recommended for their advanced immersive capabilities and integrated sensors.

\section{Control Methods For XR Prototyping}

This section seeks to answer RQ3 by reviewing the control methods explored and made available for users when interacting with virtual elements and creating XR prototypes. While traditional CAD uses a mouse and keyboard as its input, XR prototyping has employed various control methods. By control input for prototyping in XR, we mean, corresponding to Fig. \ref{workflow}, the users’ behaviour to call out and move specified virtual objects, modify a virtual object, or create objects from blank spaces. The behaviour can take place in mid-air, on touch screen surfaces, in physical-based settings (e.g., paper-based), or a combination of these.

\subsection{Controller-based Input}

A common mid-air 3D input device is the paired controller(s) with the HMDs. Recently, HMDs such as Meta Quest, Hololens and HTC Vive have employed paired controllers, including button interactions and positional tracking for input. The paired controllers of HMDs can be interpreted as a form of tangible interaction that provides an inherent and natural 3D orientation to the user, which is particularly useful for the problem of 3D data selection in volumetric data \cite{besanccon2021state}. The paired controller also benefits from advanced tracking technology, which make them more precise than self-designed marker based controllers. The technologies for positional tracking of hand controllers are not uniform but can be classified based on hardware into two categories: internal IMUs and external sensors either integrated within the HMD or deployed in the surrounding environment. Windows Mixed Reality motion controllers of Hololens and Hololens 2 obtain the position and orientation with an optical tracking sensor embedded in the HMD, which is called outside-in tracking using an external sensor to realize the tracking. Lighthouse tracking \cite{niehorster2017accuracy} adopted by HTC Vive series is also outside-in tracking with the optical sensor. The motion input of the hand controller is obtained by calculating the position and timing of the photosensors placed on the controller, which are hit by the rays emitted from the surrounding Steam Base Stations \cite{cuervo2017beyond}. In addition to precise positional tracking, paired controllers provide a larger set of easily usable and unambiguous mappable inputs compared to hand tracking or a stylus. This makes them particularly effective for interacting with a large suite of options in immersive environments or XR prototyping spaces, where clarity and flexibility in input methods are crucial.

\subsection{Pen-based Input}

From the perspective of ergonomics, pen-shaped input devices are particularly comfortable and intuitive for users due to the widespread familiarity with using pens in daily life. This makes them a popular choice for the XR 3D sketching systems. Traditionally, many articles implemented pen-shaped input devices attached with reflective markers \cite{wibowo2012dressup, jackson2016lift, arora2018symbiosissketch} or QR codes \cite{wacker2019arpen, lau2012situated,teng2017augmented}. The coordinates and thus the motion trace of the pen as input can be obtained by utilizing Camera-Only-Mapping (COM) and other passive optical motion capture techniques. Recent advancements, however, have enabled the use of pen-shaped input devices without requiring markers or additional tracking aids. These systems do not rely on semantic segmentation-based COM, but rather capture pen strokes directly to indirectly infer the motion of the pen. This approach bypasses the challenge of precisely tracking the pen tip, offering an efficient solution for 2D input tasks \cite{fender2023infinitepaint}. However, such techniques are not employed in the context of "Prototyping in XR" as discussed in this paper. The absence of markers makes it difficult to achieve accurate positional tracking in 3D space, rendering these systems unsuitable as input methods for creating 3D digital model prototypes. 
\par
Beyond optical tracking, sensor-based techniques, such as electromagnetic tracking and ultrasonic sensors, have also been utilized for mid-air input. For example, Polhemus, the 6 Degrees-Of-Freedom (DOF) electromagnetic tracking technology, is introduced for tracking the user’s hand position and orientation, and thus to obtain the user’s sketch input and to realize drawing in the air \cite{keefe2007drawing, keefe2008tech}. Tano \textit{et al.} introduced ultrasonic sensor and magnetic sensor to obtain the 3D pen input for XR prototyping, offering additional flexibility and precision in design workflows \cite{tano2013truly}.
\par
\subsection{Hand Gesture Input}
Hand gestures is a direct input method offer an intuitive and natural interaction paradigm. By bypassing the constraints of traditional input devices, gesture recognition promotes greater flexibility and immersion in XR environments, making it particularly suitable for creative tasks like drawing, sketching and modelling \cite{fechter2022comparative}.
Building on this foundation, researchers have combined hand gestures with additional input tools, such as styluses or handheld controllers, to enhance XR prototyping capabilities. For instance, Chen introduced a system that combines a tracked glove with a stylus to facilitate asymmetrical two-handed manipulation \cite{chen2011hybrid}. In Xu \textit{et al.}'s GestureSurface \cite{xu2023gesturesurface}, non-dominant hand gestures are employed as supplementary inputs for VR sketching, which validated the potential to improve the accuracy and efficiency of mid-air prototyping by providing visual cues. The Mockup Builder introduced a system of mixed gestures, incorporating both half-space input and touch input on touch displays, as will be discussed later \cite{de2012mockup}.

\subsection{Screen-based Input}

Screen-based input refers to the user using a touchscreen for inputting commands. This type of input has been extensively applied in smartphones, tablets, and other touchscreen devices. Users can manipulate the screen using their fingers or stylus pens to perform actions such as tapping, swiping, pinching, and other gestures, in order to input commands or interact with the device. 
For prototyping in XR, input on the screen surface is not limited to the interaction methods as sketching software for touch screen but was extended for specific use. 
The motion of the device in mid-air is also a core input element, used to change in the camera view \cite{xu2022applying}. Considering the scenario of holding a smartphone with one hand while performing touch screen input with the other hand, the available touch gestures are limited. The field of view (FOV) is also limited through the mobile screen. 
A straightforward approach to tackle the aforementioned challenges is to utilise the input screen exclusively as the primary controller while deploying alternative display devices characterized by an extended FOV spectrum \cite{drey2020vrsketchin}. 
Mine \textit{et al.} use the touchscreen phone as a controller instead of both controller and display \cite{mine2014making}, they build a hybrid controller that collocates a smartphone as a touch-display, a casing with physical buttons, and a microcontroller. 
\par
However, other scholars have advocated for the concurrent utilization of a singular screen device for both display and control functions. 
Mossel \textit{et al.} introduced their 3DTouch and HOMER-S \cite{mossel20133dtouch} with a multi-touch display that has been tracked full 6-DOF for the prototyping scene of rapid construction by the assembly of preset virtual model blocks. Several similar systems with different focuses \cite{marzo2014combining, dorta2016hyve, kwan2019mobi3dsketch} have been proposed to address the challenges posed by one-hand touch input, limited FOV, and lack of depth perception in prototyping in XR. These systems combine multi-touch gestures and the motion of mobile devices as inputs, aiming to provide effective solutions to these challenges.
The motion and orientation of touchscreen devices are utilized not only as inputs for camera perspective switching but also as indications for directing strokes \cite{mossel20133dtouch,lakatos2014t}. Napkin Sketch is a tablet-based AR prototyping system that uses both touchscreen devices and a stylus for input. Additionally, it incorporates a physical napkin as an intuitive, easy-to-understand interface, helping users interact with the virtual canvas and establish perspective relationships, while lowering the learning curve for new users \cite{xin2008napkin}.

When using mobile phones for mid-air prototype creation, the screen size limits users' comprehensive observation of the model, so they may have difficulty accurately grasping the size of the model. To address this issue, researchers have proposed a method that helps users more accurately create continuous strokes by varying the pitch of two tones, in order to control the position of the ``pen tip'' on a two-dimensional plane \cite{xu2022applying}. Feng \textit{et al.} enhanced the creative experience of their VR prototyping tools by providing tactile feedback through various materials on the pad's surface \cite{feng2022pressure}.

\subsection{Physical-artefacts as Inputs}

Furthermore, it should be noted that the appearance of the physical objects has also been taken as control input in the prototyping group of physical-based construction. In the lo-fi virtual scene prototyping systems with constricted creative options introduced by Hagbi \textit{et al.}, the manual drawing symbols are captured and used as control input for calling and placing the corresponding 3D model \cite{hagbi2010place, hagbi2015place, vinayak2016mobisweep}. In Eroglu \textit{et al.}'s Rilievo, designed to serve as a low-barrier creation platform for art practitioners with limited modelling expertise, a structured light scanner captures depth data of relief artworks as a supplement to the photographs, incorporating height maps as inputs into the prototyping in XR process \cite{eroglu2020rilievo}.

\section{Linking XR Prototyping to Physical Fabrication}\label{sec5}

Prototyping in XR has immense potential, driven by its intuitive interfaces, and high immersion. This technology is well-suited for digital prototyping, particularly for early-stage prototyping, which enables designers to create, manipulate, and visualize rough digital models with ease. Extending the functionality of XR by integrating prototyping in XR with real-world fabrication is the natural next step. 
This section addresses RQ4 to explore the approaches reviewed work has explored to facilitate this link to fabrication.
Our review found a relatively small number of articles that focus on linking prototyping in XR and physical manufacturing, indicating the emerging and challenging nature of this next step. We have classified them into two categories based on the fabrication methods used: manual fabrication and machine fabrication.

\subsection{Manual Fabrication}

Situated Modeling by Lau \textit{et al.} provides a simple, constrained prototyping and fabricating approach \cite{lau2012situated}. Using mark-attached handles for prototype design in AR, and in reality, using wooden blocks corresponding to different markers to build low-accuracy physical copies of virtual models. Mueller \textit{et al.} introduced their system, 'Legofy,' which converts a designed model into a LEGO-style representation to guide users in creating a low-fidelity LEGO model. This system not only simplifies the prototyping process but also generates and prints the necessary LEGO parts. Additionally, it provides assembly instructions, allowing users to manually assemble the parts. Despite the manual assembly involved, the time required for 3D printing and assembly of the LEGO model is significantly reduced compared to 3D printing the original high-fidelity model \cite{mueller2014fabrickation}. A toolkit developed by Wessely \textit{et al.} built the link by generating a cutting guide for manual fabrication from a virtual prototype \cite{wessely2018shape}. This toolkit enabled communication between physical fabrication and prototyping in XR with Blender for computers and Unity for AR devices. In the system Wiredraw, immersive guidance is provided to the user who uses the 3D squeeze pen to produce 3D wire objects, by displaying the strokes and drawing ordering the AR environment provided by the HMD \cite{yue2017wiredraw}. Hattab \textit{et al.} proposed a system to guide the manual fabrication by projecting cutting steps generated from a digital model onto material blocks in a sequential manner \cite{hattab2019rough} with the spatial augmented reality (SAR) technique. Although the system did not include a prototyping in XR process, this still presents a promising approach to turning digital models created via XR prototyping into physical handmade “body doubles”.

\subsection{Machine Fabrication}

Among the articles surveyed, 3D printing technology was the most commonly used technology for mechanized production. Integrating the function of converting the model obtained from virtual modelling into a printable model in the system is also a method. In MixFab proposed by Weichel \textit{et al.} \cite{weichel2014mixfab}, they introduced their system of prototyping in XR that generates a digital 3d model through a user’s gesture or from a scanned physical object. The 3D printable models are produced from the mesh data of the user-created model and are ready to be imported to the 3D printer manually. Similarly, Yee \textit{et al.} added an STLGenerator program in their prototyping in XR system, and thus to convert the sketch strokes into a 3D-printable object \cite{yee2009augmented}.

ROMA introduced their system including a customized Rhino plugin \cite{peng2018roma}. In this system, the 3D printing robotic arm can perform printing of the stroke or geometry that the user has just determined almost simultaneously while prototyping. More specifically, when the user sketches the prototype through the AR Head-mounted display with controllers, the spatial data of the strokes is transmitted to the Rhino plugin to build an approximate geometry, and the slicing data (printer readable execution instructions obtained from the model data) is produced and uploaded to the 3d printer arm. Their fabrication and prototyping ends communicate through serial ports.

3D printing as the representative of mechine fabrication, offers significant advantages, such as efficiency and precision. It allows for fast production based on digital models, reducing time compared to traditional methods. In systems like ROMA, user-drawn prototypes can be quickly converted into physical component for near real time printing. Additionally, it can ensure accuracy and consistency, minimizing the errors that can occur with manual fabrication. The technology also excels in creating complex shapes that are not easy to achieve manually, with systems like MixFab enabling the production of intricate models with specified features such as groove size and depth. Moreover, 3D printing can be highly automated, reducing the need for manual labor.

However, machine fabrication has limitations. 3D printing still faces material constraints, as it cannot use the wide variety of materials available in traditional manufacturing. The high initial cost of machine tools and software is another drawback, while manual methods require less investment. Furthermore, machine-fabricated products often lack the artistic touch that manual craftsmanship can offer. Lastly, 3D printing is dependent on the accuracy of digital models, and errors in model creation can result in flawed prints, whereas manual methods allow for adjustments during production.

\section{Benefits, Challenges and Future of XR Prototyping}\label{sec5}

Upon reviewing these articles and through the lens of six core building blocks of XR prototyping, we now address RQ5 by presenting the core contemporary benefits and challenges of prototyping in XR and what the future may hold for this field.
\newline
\newline
The main \textbf{benefits} of XR prototyping are:

\begin{enumerate}
    \item Increased immersion - XR prototypes can provide a more realistic experience than physical prototypes, which can help designers and engineers identify and fix problems earlier in the development process \cite{lawson2016future, van2018effectiveness, akpan2019comparative}. This immersive interaction and simulation can lead to better insights into their design and performance.
    \item Better collaboration - XR prototyping makes it easier for designers and engineers to collaborate on prototypes \cite{tano2013truly}, even if they are located in different parts of the world \cite{giunta2019investigating}. Moreover, cloud-based software platforms will allow teams to collaborate on a virtual prototype in real time.
    \item Faster design turnaround times - XR prototypes can be created and tested more quickly than physical prototypes \cite{adenauer2012virtual, nee2012augmented}, which can help reduce the time it takes to bring a product to market.
\end{enumerate}

The core \textbf{challenges of XR prototyping are}:
\begin{enumerate}
    \item Complexity in Integration - One of the major challenges in XR prototyping is the complexity of integrating XR software with existing design tools, particularly traditional CAD systems. This involves not only technical difficulties in ensuring compatibility but also the need for designers to possess both traditional design skills and expertise in advanced XR software. Accurately setting up and running simulations requires designers equipped with specialized knowledge of both traditional design principles and the XR technologies involved, which can add to the complexity.
    \item Accuracy and Fidelity of Prototypes - Achieving high accuracy and fidelity in XR prototypes remains a significant challenge. Prototypes in XR depend heavily on the quality of the design and the accuracy of the virtual environment. In addition, it is still difficult to replicate the exact properties of the real object, especially in terms of tactile feedback and highly complex material interactions. Similarly, it is difficult to fabricate prototype that retain the same appearance properties as the virtual design, such as texture and surface gloss.
    \item High Costs and Limited Accessibility -  Although XR prototyping can be less expensive than physical prototyping, high-end prototyping in XR software and hardware can be expensive, potentially out of reach for small businesses or individual designers. There is also a shortage of skilled XR developers.
\end{enumerate}

Despite the rapid development of XR technologies for prototyping and fabrication, challenges such as cross-platform compatibility, real-time feedback, and high-fidelity manufacturing still remain. In the coming years, advancements in Artificial Intelligence (AI) are expected to revolutionize the way of prototyping in XR optimized for fabrication, enabling faster, easier and more accurate design iterations. A notable breakthrough is the progress of 2D image generation models, which are evolving to enable 3D generation and control. These AI-Generated Content technologies, which can now convert the 2D photo \cite{zou2023triplane}, 2D sketch \cite{zhong2022study} or text \cite{li2023instant3d} into 3D models, are making a significant impact on AR and VR applications. These technologies have the potential to simplify the process of designing and visualizing prototypes in an immersive environment by allowing easy creation of 3D assets from simple text or novices sketches. Designers will be able to quickly prototype complex 3D structures with minimal manual effort, and even control their properties and behaviors in real-time, dramatically enhancing the XR prototyping and fabrication workflow. Future research will likely focus on enhancing the fidelity of virtual models, especially by improving material rendering, interact simulation and haptic feedback mechanisms.  As XR and fabrication technologies evolve, their integration will likely transform the way manufacturing industries approach product development, making processes faster, more cost-effective, and highly customizable. However, challenges such as ensuring cross-platform compatibility and achieving seamless integration between virtual and physical prototypes will need to be addressed before these advancements can be fully realized. In conclusion, the future of prototyping in XR and fabrication holds significant promise, as advancements in AI, machine learning, and 3D printing will likely reshape the way prototypes are designed and produced, making them faster, more efficient, and highly customizable.

\section{Conclusions}
As the landscape XR technology of continues to develop it is becoming increasingly pertinent to understand on how individuals and industries can use XR technology for artistic modelling and industrial product design. XR prototyping provides the possibility of rapid early prototyping, improved remote collaborative design efficiency, and efficient product simulation testing. However, exploration of directly linking XR prototype design with production manufacturing is still relatively limited.

We undertook a systematic review to explore the workflow of how XR prototyping is used, clarify the building blocks of XR prototype design that connect production manufacturing, and identify the potential advantages and further challenges of combining XR prototype design with production manufacturing.
A total of 54 articles related to the connection between XR prototype design and production manufacturing over the past 15 years are analyzed. Firstly, we identified the common workflows \ref{workflow} and user usage methods for XR prototype design, including physical-based construction and rendering, rapid construction by the assembly of preset virtual model blocks and prototype on a blank virtual canvas.
We summarize the theme of XR prototype design connecting production and manufacturing into six building blocks, namely control, transform, model construction and rendering, non-visual feedback, display and link to fabrication in Section\ref{topic}. We discussed the technological applications and advantages of each element in the research. Despite the challenges brought by software and hardware such as poor cross-platform compatibility, interaction delays, and difficulty in high-fidelity manufacturing, we believe that with the continuous development of intelligent systems, multimodal collaboration, and sustainable manufacturing, the integration of virtual and reality will bring more efficient, flexible, and personalized prototyping and manufacturing solutions.

\begin{appendices}

\section{}
See Table \ref{a1t}.
\begin{sidewaystable}
\caption{Range of Display Devices/Display Technologies for XR Prototyping}
\centering
\renewcommand\arraystretch{1.3}
\begin{tabular*}{\textheight}{@{\extracolsep\fill}p{0.1\textheight}p{0.2\textheight}p{0.5\textheight}p{0.1\textheight}}
\midrule
Display Devices/Display Technologies & Application & Reference & Percentage\\
\hline
\multirow{2}{*}{Screens} & Smartphone/Tablet & \cite{hagbi2010place, hagbi2015place, kim2022cor, xu2022applying, dorta2016hyve, teng2017augmented, feng2022pressure, suzuki2020realitysketch, xin2008napkin, lakatos2014t, huo2017window, mossel20133dtouch, marzo2014combining, wacker2019arpen, wacker2020heatmaps, kwan2019mobi3dsketch} & \multirow{2}{*}{43.1\%} \\
& Monitors & \cite{hagbi2010place, hagbi2015place, bergig2009place, abriata2020building, wibowo2012dressup, vinayak2016mobisweep,wessely2018shape} &\\
\hline
\multirow{3}{*}{Projection} & CAVE & \cite{jackson2016lift, chen2011hybrid, mine2014making, de2012mockup} & \multirow{3}{*}{13.7\%}\\
& Holographic & \cite{weichel2014mixfab} &\\
& Projection on other Surfaces & \cite{marner2011large,hattab2019rough} &\\
\hline
HMDs & & \cite{keefe2008tech, yee2009augmented, hagbi2010place, kim2022design, park2011ar, lau2012situated, tano2013truly, hagbi2015place, yue2017wiredraw, peng2018roma, arora2018symbiosissketch, davison2019lifebrush, xing2019hairbrush, baerentzen2019signifier, drey2020vrsketchin, horst2020bite, eroglu2020rilievo, yuan2021immersive, yu2021cassie, arora2021mid, fu2022easyvrmodeling, kim2022sketching, fechter2022comparative, zhu2022sensor, xu2023gesturesurface} & 47.1\%\\
\hline
\end{tabular*}
\end{sidewaystable}
\label{a1t}
\end{appendices}

\bmhead{Acknowledgements}
The authors acknowledge the UK's Engineering and Physical Sciences Research Council (EPSRC) for funding the Augmented Reality for Trans-Disciplinary Design of ReconFigurable Manufacturing Systems (ARTIFY) project, number 323668/0.

\section*{Declarations}

\bmhead*{Data availability}
Data are made available from the corresponding author on reasonable request.
\bmhead*{Compliance with Ethical Standards}
The authors declare that they have no conflict of interest. This study does not involve human participants and/or animals.

\bibliographystyle{agsm}
\bibliography{sn-bibliography}

@ARTICLE{Azuma2001,
  author={Azuma, R. and Baillot, Y. and Behringer, R. and Feiner, S. and Julier, S. and MacIntyre, B.},
  journal={IEEE Computer Graphics and Applications}, 
  title={Recent advances in augmented reality}, 
  year={2001},
  volume={21},
  number={6},
  pages={34-47},
  doi={10.1109/38.963459}}

@article{getuli_bim-based_2020,
	title = {{BIM}-based immersive {Virtual} {Reality} for construction workspace planning: {A} safety-oriented approach},
	volume = {114},
	doi = {10.1016/j.autcon.2020.103160},
	journal = {Automation in Construction},
	author = {Getuli, Vito and Capone, Pietro and Bruttini, Alessandro and Isaac, Shabtai},
	month = jun,
	year = {2020},
	pages = {103160}
}

@article{carmigniani2011augmented,
  title={Augmented reality technologies, systems and applications},
  author={Carmigniani, Julie and Furht, Borko and Anisetti, Marco and Ceravolo, Paolo and Damiani, Ernesto and Ivkovic, Misa},
  journal={Multimedia tools and applications},
  volume={51},
  pages={341--377},
  year={2011},
  publisher={Springer}
}

@article{mann2023extended,
  title={eXtended meta-uni-omni-Verse (XV): Introduction, Taxonomy, and State-of-the-Art},
  author={Mann, Steve and Yuan, Yu and Lamberti, Fabrizio and El Saddik, Abdulmotaleb and Thawonmas, Ruck and Prattico, Filippo Gabriele},
  journal={IEEE Consumer Electronics Magazine},
  year={2023},
  publisher={IEEE}
}

@inproceedings{cecil2016realfusion,
  title={RealFusion: An interactive workflow for repurposing real-world objects towards early-stage creative ideation},
  author={Cecil Piya, Vinayak},
  booktitle={Graphics interface},
  year={2016}
}

@inproceedings{hilliges2012holodesk,
  title={HoloDesk: direct 3d interactions with a situated see-through display},
  author={Hilliges, Otmar and Kim, David and Izadi, Shahram and Weiss, Malte and Wilson, Andrew},
  booktitle={Proceedings of the SIGCHI Conference on Human Factors in Computing Systems},
  pages={2421--2430},
  year={2012}
}

@inproceedings{weichel2014mixfab,
  title={MixFab: a mixed-reality environment for personal fabrication},
  author={Weichel, Christian and Lau, Manfred and Kim, David and Villar, Nicolas and Gellersen, Hans W},
  booktitle={Proceedings of the SIGCHI Conference on Human Factors in Computing Systems},
  pages={3855--3864},
  year={2014}
}

@inproceedings{peng2018roma,
  title={RoMA: Interactive fabrication with augmented reality and a robotic 3D printer},
  author={Peng, Huaishu and Briggs, Jimmy and Wang, Cheng-Yao and Guo, Kevin and Kider, Joseph and Mueller, Stefanie and Baudisch, Patrick and Guimbreti{\`e}re, Fran{\c{c}}ois},
  booktitle={Proceedings of the 2018 CHI conference on human factors in computing systems},
  pages={1--12},
  year={2018}
}

@inproceedings{milgram1995augmented,
  title={Augmented reality: A class of displays on the reality-virtuality continuum},
  author={Milgram, Paul and Takemura, Haruo and Utsumi, Akira and Kishino, Fumio},
  booktitle={Telemanipulator and telepresence technologies},
  volume={2351},
  pages={282--292},
  year={1995},
  organization={Spie}
}

@book{sherman2018understanding,
  title={Understanding virtual reality: Interface, application, and design},
  author={Sherman, William R and Craig, Alan B},
  year={2018},
  publisher={Morgan Kaufmann}
}

@inproceedings{nebeling2018protoar,
  title={Protoar: Rapid physical-digital prototyping of mobile augmented reality applications},
  author={Nebeling, Michael and Nebeling, Janet and Yu, Ao and Rumble, Rob},
  booktitle={Proceedings of the 2018 CHI Conference on Human Factors in Computing Systems},
  pages={1--12},
  year={2018}
}

@inproceedings{nebeling2019360proto,
  title={360proto: Making interactive virtual reality \& augmented reality prototypes from paper},
  author={Nebeling, Michael and Madier, Katy},
  booktitle={Proceedings of the 2019 CHI Conference on Human Factors in Computing Systems},
  pages={1--13},
  year={2019}
}

@inproceedings{nebeling2020xrdirector,
  title={XRDirector: A role-based collaborative immersive authoring system},
  author={Nebeling, Michael and Lewis, Katy and Chang, Yu-Cheng and Zhu, Lihan and Chung, Michelle and Wang, Piaoyang and Nebeling, Janet},
  booktitle={Proceedings of the 2020 CHI Conference on Human Factors in Computing Systems},
  pages={1--12},
  year={2020}
}

@inproceedings{gruenefeld2022vrception,
  title={VRception: Rapid Prototyping of Cross-Reality Systems in Virtual Reality},
  author={Gruenefeld, Uwe and Auda, Jonas and Mathis, Florian and Schneegass, Stefan and Khamis, Mohamed and Gugenheimer, Jan and Mayer, Sven},
  booktitle={Proceedings of the 2022 CHI Conference on Human Factors in Computing Systems},
  pages={1--15},
  year={2022}
}

@inproceedings{lau2012situated,
  title={Situated modeling: a shape-stamping interface with tangible primitives},
  author={Lau, Manfred and Hirose, Masaki and Ohgawara, Akira and Mitani, Jun and Igarashi, Takeo},
  booktitle={Proceedings of the Sixth International Conference on Tangible, Embedded and Embodied Interaction},
  pages={275--282},
  year={2012}
}

@incollection{eroglu2020rilievo,
  title={Rilievo: artistic scene authoring via interactive height map extrusion in VR},
  author={Eroglu, Sevinc and Schmitz, Patric and Martinez, Carlos Aguilera and Rusch, Jana and Kobbelt, Leif and Kuhlen, Torsten W},
  booktitle={ACM SIGGRAPH 2020 Art Gallery},
  pages={438--441},
  year={2020}
}

@article{horst2020bite,
  title={Bite-Sized Virtual Reality Learning Applications: A Pattern-Based Immersive Authoring Environment.},
  author={Horst, Robin and Naraghi-Taghi-Off, Ramtin and Rau, Linda and D{\"o}rner, Ralf},
  journal={J. Univers. Comput. Sci.},
  volume={26},
  number={8},
  pages={947--971},
  year={2020}
}

@article{kim2022cor,
  title={Cor-sketchar: Cooperative sketch-based real-time augmented reality authoring tool for crowd simulation},
  author={Kim, Gahyeon and Sung, Mankyu},
  journal={Applied Sciences},
  volume={12},
  number={15},
  pages={7416},
  year={2022},
  publisher={MDPI}
}

@article{yuan2021immersive,
  title={Immersive sketch-based tree modeling in virtual reality},
  author={Yuan, Qi and Huai, Yongjian},
  journal={Computers \& Graphics},
  volume={94},
  pages={132--143},
  year={2021},
  publisher={Elsevier}
}

@inproceedings{bergig2009place,
  title={In-place 3D sketching for authoring and augmenting mechanical systems},
  author={Bergig, Oriel and Hagbi, Nate and El-Sana, Jihad and Billinghurst, Mark},
  booktitle={2009 8th IEEE International Symposium on Mixed and Augmented Reality},
  pages={87--94},
  year={2009},
  organization={IEEE}
}

@inproceedings{hagbi2010place,
  title={In-place sketching for content authoring in augmented reality games},
  author={Hagbi, Nate and Grasset, Rapha{\"e}l and Bergig, Oriel and Billinghurst, Mark and El-Sana, Jihad},
  booktitle={2010 IEEE Virtual Reality Conference (VR)},
  pages={91--94},
  year={2010},
  organization={IEEE}
}

@article{hagbi2015place,
  title={In-place sketching for augmented reality games},
  author={Hagbi, Nate and Grasset, Raphael and Bergig, Oriel and Billinghurst, Mark and El-Sana, Jihad},
  journal={Computers in Entertainment (CIE)},
  volume={12},
  number={3},
  pages={1--18},
  year={2015},
  publisher={ACM New York, NY, USA}
}

@article{abriata2020building,
  title={Building blocks for commodity augmented reality-based molecular visualization and modeling in web browsers},
  author={Abriata, Luciano A},
  journal={PeerJ Computer Science},
  volume={6},
  pages={e260},
  year={2020},
  publisher={PeerJ Inc.}
}

@article{jackson2016lift,
  title={Lift-off: Using reference imagery and freehand sketching to create 3d models in vr},
  author={Jackson, Bret and Keefe, Daniel F},
  journal={IEEE transactions on visualization and computer graphics},
  volume={22},
  number={4},
  pages={1442--1451},
  year={2016},
  publisher={IEEE}
}

@article{cruz1992cave,
  title={The CAVE: audio visual experience automatic virtual environment},
  author={Cruz-Neira, Carolina and Sandin, Daniel J and DeFanti, Thomas A and Kenyon, Robert V and Hart, John C},
  journal={Communications of the ACM},
  volume={35},
  number={6},
  pages={64--73},
  year={1992},
  publisher={Association for Computing Machinery, Inc.}
}

@inproceedings{yu2021cassie,
  title={Cassie: Curve and surface sketching in immersive environments},
  author={Yu, Emilie and Arora, Rahul and Stanko, Tibor and B{\ae}rentzen, J Andreas and Singh, Karan and Bousseau, Adrien},
  booktitle={Proceedings of the 2021 CHI Conference on Human Factors in Computing Systems},
  pages={1--14},
  year={2021}
}

@article{park2011ar,
  title={AR-Room: a rapid prototyping framework for augmented reality applications},
  author={Park, Jong-Seung},
  journal={Multimedia tools and applications},
  volume={55},
  pages={725--746},
  year={2011},
  publisher={Springer}
}

@article{davison2019lifebrush,
  title={LifeBrush: painting, simulating, and visualizing dense biomolecular environments},
  author={Davison, Timothy and Samavati, Faramarz and Jacob, Christian},
  journal={Computers \& Graphics},
  volume={82},
  pages={232--242},
  year={2019},
  publisher={Elsevier}
}

@article{fu2022easyvrmodeling,
  title={EasyVRModeling: Easily Create 3D Models by an Immersive VR System},
  author={Fu, Zhiying and Xu, Rui and Xin, Shiqing and Chen, Shuangmin and Tu, Changhe and Yang, Chenglei and Lu, Lin},
  journal={Proceedings of the ACM on Computer Graphics and Interactive Techniques},
  volume={5},
  number={1},
  pages={1--14},
  year={2022},
  publisher={ACM New York, NY, USA}
}

@article{chen2011hybrid,
  title={A hybrid direct visual editing method for architectural massing study in virtual environments},
  author={Chen, Jian},
  journal={Collaborative Design in Virtual Environments},
  pages={131--140},
  year={2011},
  publisher={Springer}
}

@article{xu2022applying,
  title={Applying Sonification to Sketching in the Air With Mobile AR Devices},
  author={Xu, Haonan and Lyu, Fei and Huang, Jin and Tu, Huawei},
  journal={IEEE Transactions on Human-Machine Systems},
  volume={52},
  number={6},
  pages={1352--1363},
  year={2022},
  publisher={IEEE}
}

@article{kim2022sketching,
  title={Sketching in-vehicle ambient lighting in virtual reality with the Wizard-of-Oz method},
  author={Kim, Taesu and Shunayeva, Aigerim and Lee, Gyunpyo and Suk, Hyeon-Jeong},
  journal={Digital Creativity},
  volume={33},
  number={1},
  pages={49--63},
  year={2022},
  publisher={Taylor \& Francis}
}

@article{dorta2016hyve,
  title={Hyve-3D and the 3D Cursor: Architectural co-design with freedom in Virtual Reality},
  author={Dorta, Tom{\'a}s and Kinayoglu, Gokce and Hoffmann, Michael},
  journal={International Journal of Architectural Computing},
  volume={14},
  number={2},
  pages={87--102},
  year={2016},
  publisher={SAGE Publications Sage UK: London, England}
}

@article{fechter2022comparative,
  title={Comparative evaluation of WIMP and immersive natural finger interaction: A user study on CAD assembly modeling},
  author={Fechter, Marius and Schleich, Benjamin and Wartzack, Sandro},
  journal={Virtual Reality},
  volume={26},
  number={1},
  pages={143--158},
  year={2022},
  publisher={Springer}
}

@article{zhu2022sensor,
  title={Sensor-based modeling of problem-solving in virtual reality manufacturing systems},
  author={Zhu, Rui and Aqlan, Faisal and Zhao, Richard and Yang, Hui},
  journal={Expert Systems with Applications},
  volume={201},
  pages={117220},
  year={2022},
  publisher={Elsevier}
}

@article{teng2017augmented,
  title={Augmented-reality-based 3D Modeling system using tangible interface},
  author={Teng, Chin-Hung and Peng, Shueh-Shih},
  journal={Sensors and Materials},
  volume={29},
  number={11},
  pages={1545--1554},
  year={2017}
}

@article{xu2023gesturesurface,
  title={GestureSurface: VR Sketching through Assembling Scaffold Surface with Non-Dominant Hand},
  author={Xu, Xinchi and Zhou, Yang and Shao, Bingchan and Feng, Guihuan and Yu, Chun},
  journal={IEEE Transactions on Visualization and Computer Graphics},
  volume={29},
  number={5},
  pages={2499--2507},
  year={2023},
  publisher={IEEE}
}

@book{marner2011large,
  title={Large scale spatial augmented reality for design and prototyping},
  author={Marner, Michael R and Smith, Ross T and Porter, Shane R and Broecker, Markus M and Close, Benjamin and Thomas, Bruce H},
  pages={231--254},  
  year={2011},
  address		= "New {Y}ork",
  publisher={Springer}
}

@article{feng2022pressure,
  title={Pressure-sketch: a tablet-based design system in immersive VR},
  author={Feng, Shuo and He, Weiping and Wang, Shouxia and Billinghurst, Mark},
  journal={Virtual Reality},
  volume={26},
  number={3},
  pages={1207--1215},
  year={2022},
  publisher={Springer}
}

@inproceedings{baerentzen2019signifier,
  title={Signifier-Based Immersive and Interactive 3D Modeling},
  author={B{\ae}rentzen, Andreas and Frisvad, Jeppe Revall and Singh, Karan},
  booktitle={Proceedings of the 25th ACM Symposium on Virtual Reality Software and Technology},
  pages={1--5},
  year={2019}
}

@inproceedings{drey2020vrsketchin,
  title={Vrsketchin: Exploring the design space of pen and tablet interaction for 3d sketching in virtual reality},
  author={Drey, Tobias and Gugenheimer, Jan and Karlbauer, Julian and Milo, Maximilian and Rukzio, Enrico},
  booktitle={Proceedings of the 2020 CHI conference on human factors in computing systems},
  pages={1--14},
  year={2020}
}

@inproceedings{suzuki2020realitysketch,
  title={Realitysketch: Embedding responsive graphics and visualizations in AR through dynamic sketching},
  author={Suzuki, Ryo and Kazi, Rubaiat Habib and Wei, Li-Yi and DiVerdi, Stephen and Li, Wilmot and Leithinger, Daniel},
  booktitle={Proceedings of the 33rd Annual ACM Symposium on User Interface Software and Technology},
  pages={166--181},
  year={2020}
}

@inproceedings{xing2019hairbrush,
  title={Hairbrush for immersive data-driven hair modeling},
  author={Xing, Jun and Nagano, Koki and Chen, Weikai and Xu, Haotian and Wei, Li-yi and Zhao, Yajie and Lu, Jingwan and Kim, Byungmoon and Li, Hao},
  booktitle={Proceedings of the 32Nd Annual ACM Symposium on User Interface Software and Technology},
  pages={263--279},
  year={2019}
}

@inproceedings{xin2008napkin,
  title={Napkin sketch: handheld mixed reality 3D sketching},
  author={Xin, Min and Sharlin, Ehud and Sousa, Mario Costa},
  booktitle={Proceedings of the 2008 ACM symposium on Virtual reality software and technology},
  pages={223--226},
  year={2008}
}

@inproceedings{lakatos2014t,
  title={T (ether) spatially-aware handhelds, gestures and proprioception for multi-user 3D modeling and animation},
  author={Lakatos, David and Blackshaw, Matthew and Olwal, Alex and Barryte, Zachary and Perlin, Ken and Ishii, Hiroshi},
  booktitle={Proceedings of the 2nd ACM symposium on Spatial user interaction},
  pages={90--93},
  year={2014}
}

@inproceedings{mine2014making,
  title={Making VR work: building a real-world immersive modeling application in the virtual world},
  author={Mine, Mark and Yoganandan, Arun and Coffey, Dane},
  booktitle={Proceedings of the 2nd ACM symposium on Spatial user interaction},
  pages={80--89},
  year={2014}
}

@inproceedings{huo2017window,
  title={Window-shaping: 3d design ideation by creating on, borrowing from, and looking at the physical world},
  author={Huo, Ke and Vinayak and Ramani, Karthik},
  booktitle={Proceedings of the Eleventh International Conference on Tangible, Embedded, and Embodied Interaction},
  pages={37--45},
  year={2017}
}

@incollection{de2012mockup,
  title={Mockup builder: direct 3D modeling on and above the surface in a continuous interaction space},
  author={De Ara{\`u}jo, Bruno R and Casiez, G{\'e}ry and Jorge, Joaquim A},
  booktitle={Proceedings of Graphics Interface 2012},
  pages={173--180},
  year={2012}
}

@inproceedings{yee2009augmented,
  title={Augmented reality in-situ 3D sketching of physical objects},
  author={Yee, Brandon and Ning, Yuan and Lipson, Hod},
  booktitle={Intelligent UI workshop on sketch recognition},
  volume={1},
  number={2},
  year={2009},
  organization={Citeseer}
}

@inproceedings{arora2018symbiosissketch,
  title={Symbiosissketch: Combining 2d \& 3d sketching for designing detailed 3d objects in situ},
  author={Arora, Rahul and Habib Kazi, Rubaiat and Grossman, Tovi and Fitzmaurice, George and Singh, Karan},
  booktitle={Proceedings of the 2018 CHI Conference on Human Factors in Computing Systems},
  pages={1--15},
  year={2018}
}

@inproceedings{mossel20133dtouch,
  title={3DTouch and HOMER-S: intuitive manipulation techniques for one-handed handheld augmented reality},
  author={Mossel, Annette and Venditti, Benjamin and Kaufmann, Hannes},
  booktitle={Proceedings of the virtual reality international conference: laval virtual},
  pages={1--10},
  year={2013}
}

@inproceedings{marzo2014combining,
  title={Combining multi-touch input and device movement for 3D manipulations in mobile augmented reality environments},
  author={Marzo, Asier and Bossavit, Beno{\^\i}t and Hachet, Martin},
  booktitle={Proceedings of the 2nd ACM symposium on Spatial user interaction},
  pages={13--16},
  year={2014}
}

@inproceedings{wibowo2012dressup,
  title={DressUp: a 3D interface for clothing design with a physical mannequin},
  author={Wibowo, Amy and Sakamoto, Daisuke and Mitani, Jun and Igarashi, Takeo},
  booktitle={Proceedings of the Sixth International Conference on Tangible, Embedded and Embodied Interaction},
  pages={99--102},
  year={2012}
}

@article{arora2021mid,
  title={Mid-air drawing of curves on 3D surfaces in virtual reality},
  author={Arora, Rahul and Singh, Karan},
  journal={ACM Transactions on Graphics (TOG)},
  volume={40},
  number={3},
  pages={1--17},
  year={2021},
  publisher={ACM New York, NY}
}

@article{keefe2007drawing,
  title={Drawing on air: Input techniques for controlled 3D line illustration},
  author={Keefe, Daniel and Zeleznik, Robert and Laidlaw, David},
  journal={IEEE transactions on visualization and computer graphics},
  volume={13},
  number={5},
  pages={1067--1081},
  year={2007},
  publisher={IEEE}
}

@inproceedings{yue2017wiredraw,
  title={Wiredraw: 3d wire sculpturing guided with mixed reality},
  author={Yue, Ya-Ting and Zhang, Xiaolong and Yang, Yongliang and Ren, Gang and Choi, Yi-King and Wang, Wenping},
  booktitle={Proceedings of the 2017 CHI Conference on Human Factors in Computing Systems},
  pages={3693--3704},
  year={2017}
}

@inproceedings{wacker2019arpen,
  title={Arpen: Mid-air object manipulation techniques for a bimanual ar system with pen \& smartphone},
  author={Wacker, Philipp and Nowak, Oliver and Voelker, Simon and Borchers, Jan},
  booktitle={Proceedings of the 2019 CHI conference on human factors in computing systems},
  pages={1--12},
  year={2019}
}

@inproceedings{wacker2020heatmaps,
  title={Heatmaps, shadows, bubbles, rays: Comparing mid-air pen position visualizations in handheld ar},
  author={Wacker, Philipp and Wagner, Adrian and Voelker, Simon and Borchers, Jan},
  booktitle={Proceedings of the 2020 CHI Conference on Human Factors in Computing Systems},
  pages={1--11},
  year={2020}
}

@inproceedings{kwan2019mobi3dsketch,
  title={Mobi3dsketch: 3d sketching in mobile ar},
  author={Kwan, Kin Chung and Fu, Hongbo},
  booktitle={Proceedings of the 2019 CHI Conference on Human Factors in Computing Systems},
  pages={1--11},
  year={2019}
}

@inproceedings{vinayak2016mobisweep,
  title={Mobisweep: Exploring spatial design ideation using a smartphone as a hand-held reference plane},
  author={Vinayak and Ramanujan, Devarajan and Piya, Cecil and Ramani, Karthik},
  booktitle={Proceedings of the TEI'16: Tenth International Conference on Tangible, Embedded, and Embodied Interaction},
  pages={12--20},
  year={2016}
}

@inproceedings{tano2013truly,
  title={Truly useful 3D drawing system for professional designer by “life-sized and operable” feature and new interaction},
  author={Tano, Shun’ichi and Yamamoto, Shinya and Ichino, Junko and Hashiyama, Tomonori and Iwata, Mitsuru},
  booktitle={Human-Computer Interaction--INTERACT 2013: 14th IFIP TC 13 International Conference, Cape Town, South Africa, September 2-6, 2013, Proceedings, Part I 14},
  pages={37--55},
  year={2013},
  organization={Springer}
}

@article{cuervo2017beyond,
  title={Beyond reality: Head-mounted displays for mobile systems researchers},
  author={Cuervo, Eduardo},
  journal={GetMobile: Mobile Computing and Communications},
  volume={21},
  number={2},
  pages={9--15},
  year={2017},
  publisher={ACM New York, NY, USA}
}

@article{niehorster2017accuracy,
  title={The accuracy and precision of position and orientation tracking in the HTC vive virtual reality system for scientific research},
  author={Niehorster, Diederick C and Li, Li and Lappe, Markus},
  journal={i-Perception},
  volume={8},
  number={3},
  pages={2041669517708205},
  year={2017},
  publisher={Sage Publications Sage UK: London, England}
}

@inproceedings{wessely2018shape,
  title={Shape-aware material: Interactive fabrication with shapeme},
  author={Wessely, Michael and Tsandilas, Theophanis and Mackay, Wendy E},
  booktitle={Proceedings of the 31st Annual ACM Symposium on User Interface Software and Technology},
  pages={127--139},
  year={2018}
}

@inproceedings{hattab2019rough,
  title={Rough carving of 3D models with spatial augmented reality},
  author={Hattab, Ammar and Taubin, Gabriel},
  booktitle={Proceedings of the 3rd Annual ACM Symposium on Computational Fabrication},
  pages={1--10},
  year={2019}
}

@inproceedings{keefe2008tech,
  title={Tech-note: Dynamic dragging for input of 3D trajectories},
  author={Keefe, Daniel F and Zeleznik, Robert C and Laidlaw, David H},
  booktitle={2008 IEEE Symposium on 3D User Interfaces},
  pages={51--54},
  year={2008},
  organization={IEEE}
}

@inproceedings{mueller2014fabrickation,
  title={faBrickation: fast 3D printing of functional objects by integrating construction kit building blocks},
  author={Mueller, Stefanie and Mohr, Tobias and Guenther, Kerstin and Frohnhofen, Johannes and Baudisch, Patrick},
  booktitle={Proceedings of the SIGCHI Conference on Human Factors in Computing Systems},
  pages={3827--3834},
  year={2014}
}

@misc{bibAR,
  author		= "Robin, Seiler",
  title			= "Microsoft brings Windows 11 to HoloLens 2", 
  year			= "2023",
  note			= "\url{https://blogs.windows.com/windowsexperience/2023/04/13/microsoft-brings-windows-11-to-hololens-2/}"
}

@misc{poke,
  author		= "{{The Pokémon GO team}}",
  title			= "Pokémon GO AR Photo Contest Rules", 
  year			= "2017",
  note			= "\url{https://pokemongolive.com/en/post/arphotocontest/?hl=zh_hant}"
}

@misc{varjo,
  author		= "{VARJO}",
  title			= "Highest Resolution Virtual Reality headset for Professional", 
  year			= "2023",
  note			=  {Accessed: 3rd April 2025},
    howpublished = {\url{https://varjo.com/products/varjo-vr-3/}}
}

@misc{cave,
  author		= "{{Visbox, Inc.}}",
  title			= "VisCube™ M4, M5 | CAVE Immersive 3D Display", 
  year			= "2020",
  note			= {Accessed: 3rd April 2025},
howpublished = {\url{http://www.visbox.com/products/cave/viscube-m4/}}
}

@misc{honda,
  author		= "{{Honda News}}",
  title			= "New Honda Design Video: How the Latest VR Technology Is Accelerating the Design Process of Honda EVs", 
  year			= "2023",
  note			= {Accessed: 3rd April 2025},
howpublished = {\url{https://hondanews.com/en-US/releases/release-4e58b4e0fcd795affa5685a66a2a3166}}
}

@misc{ford,
  author		= "{{Ford Media Center}}",
  title			= "FORD COLLABORATION WITH GRAVITY SKETCH INTRODUCES CO-CREATION FEATURE, ALLOWING DESIGNERS ACROSS GLOBE TO WORK IN SAME VIRTUAL REALITY SPACE", 
  year			= "2019",
  note			= {Accessed: 3rd April 2025},
howtopublished = {\url{https://media.ford.com/content/fordmedia/fna/us/en/news/2019/05/06/ford-collaboration-gravity-sketch-co-creation.html}}
}

@misc{anta,
author = "{{Gravity Sketch}}",
year = "2021",
  title			= "Five questions on {VR} design with {Sean O’Shea} from {Anta}", 
  howpublished = {\url{https://tinyurl.com/y2wehmb9}},
note = {Accessed: 2023-10-13}
}

@misc{GoogleBlocks,
  author		= "{{XR for designer}}",
  title			= "Rapid VR prototyping without coding in 2019", 
  year			= "2019",
  howtopublished			= {\url{https://tinyurl.com/22dhebh6}},
note = {Accessed: 2023-10-13}

}

@misc{VRcompare,
    author		= "{VRcompare}",
    title = "VRcompare - The Internet's Largest VR and AR Headset Database",
    howtopublished = {\url{https://vr-compare.com/}},
    year			= "2024",
    note = {Accessed: 2024-05-13}
}

@misc{gamingmonitor,
    title = "Gaming Monitor Comparison - May 2024",
    author		= "{TopChoice}", 
    howtopublished = {\url{https://www.topchoice.co.uk/gaming-monitor}},
    year			= "2024",
    note = {Accessed: 2024-05-13}
}

@inproceedings{freitas2020systematic,
  title={A systematic review of rapid prototyping tools for augmented reality},
  author={Freitas, Gabriel and Pinho, Marcio Sarroglia and Silveira, Milene Selbach and Maurer, Frank},
  booktitle={2020 22nd Symposium on Virtual and Augmented Reality (SVR)},
  pages={199--209},
  year={2020},
  organization={IEEE}
}

@article{schiavi2022bim,
  title={BIM data flow architecture with AR/VR technologies: Use cases in architecture, engineering and construction},
  author={Schiavi, Barbara and Havard, Vincent and Beddiar, Karim and Baudry, David},
  journal={Automation in Construction},
  volume={134},
  pages={104054},
  year={2022},
  publisher={Elsevier}
}

@article{kordass2002virtual,
  title={The virtual articulator in dentistry: concept and development},
  author={Korda{\ss}, Bernd and G{\"a}rtner, Christian and S{\"o}hnel, Andreas and Bisler, Alexander and Vo{\ss}, Gerrit and Bockholt, Ulrich and Seipel, Stefan},
  journal={Dental Clinics},
  volume={46},
  number={3},
  pages={493--506},
  year={2002},
  publisher={Elsevier}
}

@article{monaghesh2023application,
  title={Application of virtual reality in dental implants: a systematic review},
  author={Monaghesh, Elham and Negahdari, Ramin and Samad-Soltani, Taha},
  journal={BMC Oral Health},
  volume={23},
  number={1},
  pages={603},
  year={2023},
  publisher={Springer}
}

@article{moerland2021application,
  title={Application of VR technology in the aircraft cabin design process},
  author={Moerland-Masic, Ivana and Reimer, Fabian and Bock, Thomas M and Meller, Frank and Nagel, Bj{\"o}rn},
  journal={CEAS Aeronautical Journal},
  pages={1--10},
  year={2021},
  publisher={Springer}
}

@article{akpan2019comparative,
  title={A comparative evaluation of the effectiveness of virtual reality, 3D visualization and 2D visual interactive simulation: an exploratory meta-analysis},
  author={Akpan, Ikpe Justice and Shanker, Murali},
  journal={Simulation},
  volume={95},
  number={2},
  pages={145--170},
  year={2019},
  publisher={SAGE Publications Sage UK: London, England}
}

@incollection{adenauer2012virtual,
  title={Virtual reality technologies for creative design},
  author={Adenauer, Julian and Israel, Johann Habakuk and Stark, Rainer},
  booktitle={CIRP Design 2012: Sustainable Product Development},
  pages={125--135},
  year={2012},
  publisher={Springer}
}

@article{lawson2016future,
  title={Future directions for the development of virtual reality within an automotive manufacturer},
  author={Lawson, Glyn and Salanitri, Davide and Waterfield, Brian},
  journal={Applied ergonomics},
  volume={53},
  pages={323--330},
  year={2016},
  publisher={Elsevier}
}

@inproceedings{van2018effectiveness,
  title={Effectiveness of virtual reality in participatory urban planning: A case study},
  author={Van Leeuwen, Jos P and Hermans, Klaske and Jylh{\"a}, Antti and Quanjer, Arnold Jan and Nijman, Hanke},
  booktitle={Proceedings of the 4th Media Architecture Biennale Conference},
  pages={128--136},
  year={2018}
}

@article{nee2012augmented,
  title={Augmented reality applications in design and manufacturing},
  author={Nee, Andrew YC and Ong, SK and Chryssolouris, George and Mourtzis, Dimitris},
  journal={CIRP annals},
  volume={61},
  number={2},
  pages={657--679},
  year={2012},
  publisher={Elsevier}
}

@inproceedings{giunta2019investigating,
  title={Investigating the impact of spatial augmented reality on communication between design session participants-A pilot study},
  author={Giunta, Lorenzo and Guefrache, Fatma Ben and Dekoninck, Elies and Gopsill, James and O'Hare, Jamie and Morosi, Federico},
  booktitle={Proceedings of the Design Society: International Conference on Engineering Design},
  volume={1},
  number={1},
  pages={1973--1982},
  year={2019},
  organization={Cambridge University Press}
}

@article{chang2020virtual,
  title={Virtual reality sickness: a review of causes and measurements},
  author={Chang, Eunhee and Kim, Hyun Taek and Yoo, Byounghyun},
  journal={International Journal of Human--Computer Interaction},
  volume={36},
  number={17},
  pages={1658--1682},
  year={2020},
  publisher={Taylor \& Francis}
}

@article{saredakis2020factors,
  title={Factors associated with virtual reality sickness in head-mounted displays: a systematic review and meta-analysis},
  author={Saredakis, Dimitrios and Szpak, Ancret and Birckhead, Brandon and Keage, Hannah AD and Rizzo, Albert and Loetscher, Tobias},
  journal={Frontiers in human neuroscience},
  volume={14},
  pages={96},
  year={2020},
  publisher={Frontiers Media SA}
}

@article{el2019survey,
  title={Survey on depth perception in head mounted displays: distance estimation in virtual reality, augmented reality, and mixed reality},
  author={El Jamiy, Fatima and Marsh, Ronald},
  journal={IET Image Processing},
  volume={13},
  number={5},
  pages={707--712},
  year={2019},
  publisher={Wiley Online Library}
}

@article{mcgill2022augmented,
  title={Augmented, Virtual and Mixed Reality Passenger Experiences},
  author={McGill, Mark and Li, Gang and Ng, Alex and Bajorunaite, Laura and Williamson, Julie and Pollick, Frank and Brewster, Stephen},
  journal={User Experience Design in the Era of Automated Driving},
  pages={445--475},
  year={2022},
  publisher={Springer}
}

@article{jimeno2007state,
  title={State of the art of the virtual reality applied to design and manufacturing processes},
  author={Jimeno, Antonio and Puerta, Alberto},
  journal={The International Journal of Advanced Manufacturing Technology},
  volume={33},
  pages={866--874},
  year={2007},
  publisher={Springer}
}

@misc{OpenXR,
    author		= "{Blender}",
    title = "Developer - Doc - Feature - Virtual reality - OpenXR",
    note = "\url{https://developer.blender.org/docs/features/gpu/viewports/xr/}",
    year			= "2024",
}

@misc{hpg2,
  author       = {HP},
  title        = {HP Reverb G2 Omnicept Edition},
  year         = {2019},
  note          = {\url{https://h20195.www2.hp.com/v2/GetDocument.aspx?docname=c06699581}}
}

@article{tyndall2010aacods,
  title={AACODS checklist},
  author={Tyndall, Jess},
  year={2010},
  publisher={< bound method Organization. get\_name\_with\_acronym of< Organization~…}
}

@article{lorenz2016cad,
  title={CAD to VR--a methodology for the automated conversion of kinematic CAD models to virtual reality},
  author={Lorenz, Mario and Spranger, Michael and Riedel, Tino and P{\"u}rzel, Franziska and Wittstock, Volker and Klimant, Philipp},
  journal={Procedia Cirp},
  volume={41},
  pages={358--363},
  year={2016},
  publisher={Elsevier}
}

@article{kim2022design,
  title={Design and implementation of opc ua-based vr/ar collaboration model using cps server for vr engineering process},
  author={Kim, Jeehyeong and Jeong, Jongpil},
  journal={Applied Sciences},
  volume={12},
  number={15},
  pages={7534},
  year={2022},
  publisher={MDPI}
}

@article{bai2024universal,
  title={Universal facial encoding of codec avatars from {VR} headsets},
  author={Bai, Shaojie and Wang, Te-Li and Li, Chenghui and Venkatesh, Akshay and Simon, Tomas and Cao, Chen and Schwartz, Gabriel and Wrench, Ryan and Saragih, Jason and Sheikh, Yaser and others},
  journal={arXiv preprint arXiv:2407.13038},
  year={2024}
}

@inproceedings{besanccon2021state,
  title={The state of the art of spatial interfaces for 3D visualization},
  author={Besan{\c{c}}on, Lonni and Ynnerman, Anders and Keefe, Daniel F and Yu, Lingyun and Isenberg, Tobias},
  booktitle={Computer Graphics Forum},
  volume={40},
  number={1},
  pages={293--326},
  year={2021},
  organization={Wiley Online Library}
}

@inproceedings{fender2023infinitepaint,
  title={InfinitePaint: Painting in Virtual Reality with Passive Haptics Using Wet Brushes and a Physical Proxy Canvas},
  author={Fender, Andreas Rene and Roberts, Thomas and Luong, Tiffany and Holz, Christian},
  booktitle={Proceedings of the 2023 CHI Conference on Human Factors in Computing Systems},
  pages={1--13},
  year={2023}
}

@misc{ARK,
  author       = {Apple},
  title        = {Index of /artoolkit},
  year         = {2021},
  note          = {\url{https://developer.apple.com/documentation/arkit/}}}

@misc{ARC,
  author       = {Google},
  title        = {Index of /artoolkit},
  year         = {2018},
  note          = {\url{https://developers.google.com/ar/}}
}

@misc{GSsneaker,
  author       = {Joey Khamis},
  title        = {INSTINCT},
  year         = {2022},
note = {Accessed: 3rd April 2025},
  howtopublished          = {\url{https://www.khamisstudio.com/gravitysketch}}
}

@misc{GGBlocks,
  author       = {Google},
  title        = {Blocks: Easily Create 3D Objects in VR},
  year         = {2017},
  note          = {\url{https://www.khamisstudio.com/gravitysketch}}
}

@misc{gravityS,
  author       = {Gravity Sketch},
  title        = {Gravity Sketch 6.0: putting the right tools at your fingertips, right when you need them},
  year         = {2023},
  note          = {\url{https://gravitysketch.com/blog-post/updates/gravity-sketch-6-0/}}
}

@misc{tiltB,
  author       = {Google},
  title        = {Tilt Brush},
  year         = {2016},
  note          = {\url{https://support.google.com/tiltbrush/answer/6389710?hl=en}}
}

@misc{opB,
  author       = {Open Brush},
  title        = {Open Brush},
  year         = {2020},
  note          = {\url{https://openbrush.app/}}
}

@misc{MicroMa,
  author       = {Microsoft Marquette},
  title        = {Microsoft Maquette Beta overview - Mixed Reality | Microsoft Learn},
  year         = {2018},
  note          = {\url{https://learn.microsoft.com/en-us/windows/mixed-reality/design/maquette}}
}

@article{li2023instant3d,
  title={Instant3d: Fast text-to-3d with sparse-view generation and large reconstruction model},
  author={Li, Jiahao and Tan, Hao and Zhang, Kai and Xu, Zexiang and Luan, Fujun and Xu, Yinghao and Hong, Yicong and Sunkavalli, Kalyan and Shakhnarovich, Greg and Bi, Sai},
  journal={arXiv preprint arXiv:2311.06214},
  year={2023}
}

@article{zou2023triplane,
  title={Triplane Meets Gaussian Splatting: Fast and Generalizable Single-View 3D Reconstruction with Transformers},
  author={Zou, Zi-Xin and Yu, Zhipeng and Guo, Yuan-Chen and Li, Yangguang and Liang, Ding and Cao, Yan-Pei and Zhang, Song-Hai},
  journal={arXiv preprint arXiv:2312.09147},
  year={2023}
}

@article{zhong2022study,
  title={A study of deep single sketch-based modeling: View/style invariance, sparsity and latent space disentanglement},
  author={Zhong, Yue and Gryaditskaya, Yulia and Zhang, Honggang and Song, Yi-Zhe},
  journal={Computers \& Graphics},
  volume={106},
  pages={237--247},
  year={2022},
  publisher={Elsevier}
}

\end{document}